\documentclass[linenumbers,twocolumn]{aastex631}
\usepackage{lineno}
\linenumbers
\usepackage{upgreek}
\usepackage{gensymb}
\usepackage{graphicx}
\usepackage{subfigure}
\usepackage{enumerate}
\usepackage{booktabs}
\usepackage{amsmath}
\usepackage{kotex}

\usepackage{scalerel}
\newcommand\HII{H\protect\scaleto{$II$\ }{1.2ex}}

\begin{document}
\nolinenumbers

\title{The JCMT BISTRO Survey: Unveiling the Magnetic Fields around Galactic Center}

\author[0009-0003-5699-2723]{Meng-Zhe Yang}
\affiliation{Institute of Astronomy and Department of Physics, National Tsing Hua University, Hsinchu 30013, Taiwan}

\correspondingauthor{Meng-Zhe Yang}
\email{jeremys93102@gapp.nthu.edu.tw}

\author[0000-0001-5522-486X]{Shih-Ping Lai}
\affiliation{Institute of Astronomy and Department of Physics, National Tsing Hua University, Hsinchu 30013, Taiwan}
\affiliation{Academia Sinica Institute of Astronomy and Astrophysics, No.1, Sec. 4., Roosevelt Road, Taipei 10617, Taiwan}

\author[0000-0001-5996-3600]{Janik Karoly}
\affiliation{Department of Physics and Astronomy, University College London, WC1E 6BT London, UK}

\author[0000-0002-8557-3582]{Kate Pattle}
\affiliation{Department of Physics and Astronomy, University College London, WC1E 6BT London, UK}

\author[0000-0003-2619-9305]{Xing Lu (呂行)} %
\affiliation{Shanghai Astronomical Observatory, Chinese Academy of Sciences, 80 Nandan Road, Shanghai 200030, People’s Republic of China}

\author{David Eden}
\affiliation{Department of Physics, University of Bath, Claverton Down, Bath BA2 7AY, UK}
\affiliation{Armagh Observatory and Planetarium, College Hill, Armagh BT61 9DB, UK}

\author[0000-0002-6868-4483]{Sheng-Jun Lin}
\affiliation{Academia Sinica Institute of Astronomy and Astrophysics, No.1, Sec. 4., Roosevelt Road, Taipei 10617, Taiwan}

\author[0000-0002-5391-5568]{Frédérick Poidevin}
\affiliation{Instituto de Astrofísica de Canarias, Vía Láctea, 38205 La Laguna, Tenerife, Spain.}
\affiliation{Universidad de La Laguna, Departamento de Astrofísica,  38206 La Laguna, Tenerife, Spain.}

\author[0000-0002-4541-0607]{Ekta Sharma}
\affiliation{CAS Key Laboratory of FAST, National Astronomical Observatories, Chinese Academy of Sciences, People's Republic of China}

\author[0000-0001-7866-2686]{Jihye Hwang}
\affiliation{Korea Astronomy and Space Science Institute, 776 Daedeokdae-ro, Yuseong-gu, Daejeon 34055, Republic of Korea}

\author[0000-0001-9930-9240]{Lapo Fanciullo}
\affiliation{National Chung Hsing University, 145 Xingda Rd., South Dist., Taichung City 402, Taiwan}

\author[0000-0001-8749-1436]{Mehrnoosh Tahani}
\affiliation{Banting and KIPAC Fellow: Kavli Institute for Particle Astrophysics and Cosmology (KIPAC), Stanford University, Stanford, CA, United States}

\author[0000-0003-2777-5861]{Patrick M. Koch}
\affiliation{Academia Sinica Institute of Astronomy and Astrophysics, No.1, Sec. 4., Roosevelt Road, Taipei 10617, Taiwan}

\author[0000-0003-4366-6518]{Shu-ichiro Inutsuka}
\affiliation{Department of Physics, Graduate School of Science, Nagoya University, Furo-cho, Chikusa-ku, Nagoya 464-8602, Japan}

\author{Valentin J. M. Le Gouellec}
\affiliation{SOFIA Science Center, Universities Space Research Association, NASA Ames Research Center, Moffett Field, California 94035, USA}
\affiliation{Université Paris-Saclay, CNRS, CEA, Astrophysique, Instrumentation et Modélisation de Paris-Saclay, 91191 Gif-sur-Yvette, France}

\author[0000-0002-7022-4742]{Hao-Yuan Duan}
\affiliation{Taipei Astronomical Museum, Taipei, Taiwan}
\affiliation{Institute of Astronomy and Department of Physics, National Tsing Hua University, Hsinchu 30013, Taiwan11}

\author[0000-0002-6668-974X]{Jia-Wei Wang}
\affiliation{Academia Sinica Institute of Astronomy and Astrophysics, No.1, Sec. 4., Roosevelt Road, Taipei 10617, Taiwan}

\author{Gary Fuller}
\affiliation{Jodrell Bank Centre for Astrophysics, School of Physics and Astronomy, University of Manchester, Oxford Road, Manchester, M13 9PL, UK}

\author{Ray S. Furuya}
\affiliation{Institute of Liberal Arts and Sciences Tokushima University, Minami Jousanajima-machi 1-1, Tokushima 770-8502, Japan}

\author{Qilao Gu}
\affiliation{Shanghai Astronomical Observatory, Chinese Academy of Sciences, 80 Nandan Road, Shanghai 200030, People’s Republic of China}

\author[0000-0003-1853-0184]{Tetsuo Hasegawa (長谷川哲夫)} %
\affiliation{National Astronomical Observatory of Japan, National Institutes of Natural Sciences, Osawa, Mitaka, Tokyo 181-8588, Japan}

\author{Guangxing Li}
\affiliation{Yunnan University}

\author[0000-0002-4774-2998]{Junhao Liu (劉峻豪)} %
\affiliation{Division of ALMA, National Astronomical Observatory of Japan, Mitaka, Tokyo 181-8588, Japan}

\author[0000-0001-6258-7474]{M. S. Akshaya}
\affiliation{Korea Astronomy and Space Science Institute, 776 Daedeokdae-ro, Yuseong-gu, Daejeon 34055, Republic of Korea}

\author[0000-0002-6398-7530]{Bijas Najimudeen}
\affiliation{Jodrell Bank Centre for Astrophysics, School of Physics and Astronomy, University of Manchester, Oxford Road, Manchester, M13 9PL, UK}

\author[0000-0002-6488-8227]{Le Ngoc Tram}
\affiliation{University of Science and Technology of Hanoi, Vietnam Academy of Science and Technology, 18 Hoang Quoc Viet, Hanoi, Vietnam}

\author[0000-0003-1140-2761]{Derek Ward-Thompson}
\affiliation{Jeremiah Horrocks Institute, University of Central Lancashire, Preston PR1 2HE, UK}

\author[0000-0002-1959-7201]{Doris Arzoumanian}
\affiliation{Division of Science, National Astronomical Observatory of Japan, 2-21-1 Osawa, Mitaka, Tokyo 181-8588, Japan}

\author[0000-0002-9289-2450]{James Di Francesco}
\affiliation{NRC Herzberg Astronomy and Astrophysics, 5071 West Saanich Road, Victoria, BC V9E 2E7, Canada}
\affiliation{Department of Physics and Astronomy, University of Victoria, Victoria, BC V8W 2Y2, Canada}

\author[0000-0001-8746-6548]{Yasuo Doi}
\affiliation{Department of Earth Science and Astronomy, Graduate School of Arts and Sciences, The University of Tokyo, 3-8-1 Komaba, Meguro, Tokyo 153-8902, Japan}

\author[0000-0003-2017-0982]{Thiem Hoang}
\affiliation{Korea Astronomy and Space Science Institute, 776 Daedeokdae-ro, Yuseong-gu, Daejeon 34055, Republic of Korea}

\author[0000-0001-7379-6263]{Ji-hyun Kang}
\affiliation{Korea Astronomy and Space Science Institute, 776 Daedeokdae-ro, Yuseong-gu, Daejeon 34055, Republic of Korea}

\author[0000-0003-2815-7774]{Jungmi Kwon}
\affiliation{Department of Astronomy, University of Tokyo, 7-3-1 Hongo, Bunkyo-ku, Tokyo 113-0033, Japan}

\author[0000-0003-4022-4132]{Woojin Kwon}
\affiliation{Department of Earth Science Education, Seoul National University, 1 Gwanak-ro, Gwanak-gu, Seoul 08826, Republic of Korea}
\affiliation{SNU Astronomy Research Center, Seoul National University, 1 Gwanak-ro, Gwanak-gu, Seoul 08826, Republic of Korea}
\affiliation{The Center for Educational Research, Seoul National University, 1 Gwanak-ro, Gwanak-gu, Seoul 08826, Republic of Korea}

\author[0000-0002-3179-6334]{Chang Won Lee (이창원)} %
\affiliation{Korea Astronomy and Space Science Institute, 776 Daedeokdae-ro, Yuseong-gu, Daejeon 34055, Republic of Korea}

\author[0000-0002-5286-2564]{Tie Liux (이창원)} %
\affiliation{Key Laboratory for Research in Galaxies and Cosmology, Shanghai Astronomical Observatory, Chinese Academy of Sciences, 80 Nandan Road, Shanghai 200030, People’s Republic of China}

\author[0000-0002-8234-6747]{Takashi Onaka}
\affiliation{Department of Astronomy, Graduate School of Science, The University of Tokyo, 7-3-1 Hongo, Bunkyo-ku, Tokyo 113-0033, Japan}

\author{Sarah Sadavoy}
\affiliation{Department for Physics, Engineering Physics and Astrophysics, Queen's University, Kingston, ON, K7L 3N6, Canada}

\author[0000-0002-6510-0681]{Motohide Tamura}
\affiliation{National Astronomical Observatory of Japan, National Institutes of Natural Sciences, Osawa, Mitaka, Tokyo 181-8588, Japan}
\affiliation{Department of Astronomy, Graduate School of Science, The University of Tokyo, 7-3-1 Hongo, Bunkyo-ku, Tokyo 113-0033, Japan}

\author[0000-0002-0794-3859]{Pierre Bastien}
\affiliation{Centre de recherche en astrophysique du Québec \& département de physique, Université de Montréal, C.P. 6128 Succ. Centre-ville, Montréal, QC, H3C 3J7, Canada}

\author[0000-0001-6524-2447]{David Berry}
\affiliation{East Asian Observatory, 660 N. A'oh\={o}k\={u} Place, University Park, Hilo, HI 96720, USA}

\author[0000-0002-0859-0805]{Simon Coudé}
\affiliation{SOFIA Science Center, Universities Space Research Association, NASA Ames Research Center, Moffett Field, California 94035, USA}

\author[0000-0002-5093-5088]{Keping Qiu}
\affiliation{School of Astronomy and Space Science, Nanjing University, 163 Xianlin Avenue, Nanjing 210023, People's Republic of China}
\affiliation{Key Laboratory of Modern Astronomy and Astrophysics (Nanjing University), Ministry of Education, Nanjing 210023, People's Republic of China}

\begin{abstract}
\nolinenumbers
We acquired 450\,$\rm \upmu$m and 850 $\rm \upmu$m dust continuum polarization observations toward the inner region of the Central Molecular Zone (CMZ) as part of the B-Fields In Star-Forming Region Observations (BISTRO) survey using the POL-2 polarimeter on the James Clerk Maxwell Telescope.  These observations encompassed three dense structures: the 20 km\,s$\rm ^{-1}$ cloud (20MC), 50 km\,s$\rm ^{-1}$ cloud (50MC), and circumnuclear disk (CND).
Our aim is to investigate the magnetic field morphology and strength in the inner region of the CMZ using polarized dust continuum and the Davis-Chandrasekhar-Fermi  method.
The magnetic field morphology is highly ordered in all three dense regions.
The plane-of-sky magnetic field strengths are $\sim$1\,mG for the 20MC and the 50MC, and $\sim$2\,mG for the CND.
We compare the energy contributions of turbulence, gravity, and thermal motion with that of the magnetic field using the plasma $\beta$, mass-to-flux ratio, and Alfvén Mach number.
The outcomes reveal the magnetic field stands out as the predominant factor within the inner region of the CMZ.
The dominance of the magnetic field may explain the low star-forming rate in the CMZ.
We further investigate the dust grain alignment efficiency by exploring the relationship between polarization fraction and total intensity. 
The results suggest that dust grains are well aligned with the magnetic fields.

\end{abstract}

\keywords{submillimeter: ISM --- ISM: magnetic fields --- ISM: clouds --- stars: formation --- techniques:polarimetric}


\section{Introduction} \label{sec:intro}

The Central Molecular Zone (CMZ), the inner region of the Milky Way with a radius of $\sim200\,\text{pc}$ \citep{1996ARA&A..34..645M}, contains over $10^7\ M_{\odot}$ of molecular gas \citep{2007A&A...467..611F,2017ApJ...839....1L}.
The molecular clouds within the CMZ are warmer and more turbulent compared to those in the galactic disk \citep[][]{2012PASJ...64..111T}.
The star formation rate (SFR) is typically proportional to the gas surface density or the gas volume density \citep[][]{1959ApJ...129..243S,1998ApJ...498..541K,2012ARA&A..50..531K}.
The typical SFR in the CMZ, however, is approximately 10 times lower than that in other regions of the Milky Way \citep[][]{2013MNRAS.429..987L, 2014MNRAS.440.3370K}.

\citet{2019ApJ...872..171L} investigated the SFR in six massive CMZ clouds, including the 20\,km\,s$^{-1}$ cloud, the 50\,km\,s$^{-1}$ cloud, G.0253+0.016 (the Brick), Sgr\,B1-off, and Sgr\,C, using observations from the SMA and VLA, also utilizing data from \citet{2018ApJ...853..171G} for Sgr\,B2.
Among these regions, Sgr\,B2 and Sgr\,C follow the dense gas star formation relation inferred from nearby clouds, however, the SFRs in the other four clouds are approximately one order of magnitude lower.
\citet{2019ApJ...872..171L} suggested that the low SFR is attributed to the presence of strong turbulence, resulting in a significantly lower amount of gas confined in gravitationally bound cores, indicating a magnetically supercritical environment.
Apart from these massive molecular clouds, in the region close to Sgr A*, turbulence dominates the internal energy and suppresses star formation at distances greater than 1 pc. 
Meanwhile, tidal forces dominate within 1 pc, as suggested by the un-magnetized virial theorem \citep[][]{2021ApJ...913...94H}.

In general, the initiation of the star-forming process occurs within magnetized and turbulent molecular clouds.
The magnetic field can play multiple roles, inhibiting gravitational collapse, reducing star formation efficiency \citep[e.g.][]{2008ApJ...687..354N, 2019ApJ...876...42W}, and mitigating angular momentum \citep[e.g.,][]{1985prpl.conf..320M, 1986ApJ...308..781M}.
Characterizing the magnetic field within the CMZ is therefore pivotal for understanding the underlying reasons for its relatively low SFR.

The 20\,km\,s$^{-1}$ cloud and 50\,km\,s$^{-1}$ cloud, which are named after their local standard of the rest radial velocities (hereinafter referred to 20MC and 50MC, respectively), are two massive molecular clouds close to Sgr\,A* with low SFRs, as mentioned above.
For the 20MC, \citet{2017IAUS..322...99L} proposed that the star forming process is still in an early evolutionary phase based on the observable $\rm H_2O$ masers combined with the non-detection of Class II $\rm CH_3OH$ masers and ultra-compact \HII region.
\citet{2015MNRAS.447.1059K} proposed that star formation within this massive cloud is possibly triggered by tidal compression.
In contrast, several compact \HII regions have been observed in the 50MC \citep[e.g.,][]{2010ApJ...725.1429Y, 2011ApJ...735...84M}.
This cloud, with huge amounts of shocked molecular gas traced by SiO (2-1), was proposed to be a candidate of a massive star-forming region induced by cloud-cloud collision \citep[][]{2015ASPC..499..249T, 2019ApJ...872..121U}.
Both tidal compression and cloud-cloud collision can lead to severe turbulence, which may play a crucial role during the star forming process.

In addition to severe turbulence, the magnetic field also appears to be strong in CMZ clouds \citep[e.g.,][]{1989NuPhS..10...59M}.
The magnetic field orientations are highly ordered in this region \citep[e.g.][]{2018ApJ...862..150H,2019A&A...630A..74M,2022MNRAS.511..829H}.
Recently, \citet{2024ApJ...969..150P} presented the magnetic field morphology within the CMZ region using SOFIA/HAWC+ observations at 214\,$\rm \upmu m$. 
The distribution of polarization pseudovectors reveals a bimodal pattern in the CMZ magnetic field orientations, showing field components that align either parallel or perpendicular to the Galactic plane.

For the active star-forming cloud Sgr\,B2, \citet{1999ApJ...520..706C} measured a line-of-sight (LOS) magnetic field strength of 480\,$\rm \upmu G$ using Zeeman measurements and estimated the mass-to-flux ratio to be $\sim$2.6, suggesting the cloud is magnetically supercritical.
In comparison, \citet{2015ApJ...799...74P} estimated the plane-of-sky (POS) magnetic field strength within the inactive star-forming cloud G0.253+0.016 using the Davis-Chandrasekhar-Fermi method with 850\,$\rm \upmu m$ polarization observations taken by the JCMT equipped with the polarimeter SCUPOL. 
They obtained a POS magnetic field strength of $\sim$5\,mG and estimated the mass-to-flux ratio to be around 0.6, suggesting that the magnetic fields may impede gravitational collapse.
These findings suggest the magnetic field may be an important factor in determining the SFR in the CMZ.

The closest molecular gas reservoir near the Galactic Center is the circumnuclear disk (CND) of the Milky Way, which is a torus-like structure rotating around Sgr\,A* with a radius from 2\,pc to 5\,pc \citep[e.g.,][]{1999ApJ...511..761L,2018PASJ...70...85T,2019ApJ...885L..20H,2024ApJ...969..150P}.
The magnetic field strength towards the CND was measured to be $\sim3$\,mG, as obtained from Zeeman measurements by \citet{1995ApJ...445L.113P}.
The magnetic field is dynamically important and influencing the gas motion in CND.
It not only supports the gas against gravity but also contributes to accretion toward Sgr\,A* by removing angular momentum.
\citet{1990ApJ...362..120W} modeled the magnetic field in this region as an accretion disk threaded by open magnetic field lines. 
The angular momentum is effectively removed when the angle between the poloidal component of the magnetic field (the component in the $r$-$z$ plane, where $r$ is the radial direction and $z$ is the symmetry axis relative to a reference torus) and the disk surface is less than $\sim$\,55$\degree$.
\citet{1990ApJ...362..114H,1993ApJ...417..565H} measured the polarization of the far-infrared thermal emission in the CND, and showed that the observations were consistent with the model proposed by \citet{1990ApJ...362..120W}.

In this paper, we present the 450$\,\upmu m$ and 850$\,\upmu m$ observations towards the Galactic Center taken by the Submillimetre Common-User Bolometer Array 2 \citep[SCUBA-2,][]{2013MNRAS.430.2513H} instrument and its associated polarimeter POL-2 \citep[][]{2011ASPC..449...68B,2016SPIE.9914E..03F} equipped on the James Clerk Maxwell Telescope (JCMT).
These data were taken as part of the B-fields In STar forming Regions Observations (BISTRO) survey \citep[][]{2017ApJ...842...66W,2018ApJ...859....4K,2018ApJ...861...65S,2019ApJ...877...43L}.
We explore the magnetic field morphology and strengths within the 20MC, 50MC, and CND.
Meanwhile, an overview of the survey data and a global analysis of the 850\,$\upmu$m magnetic field across the entire CMZ are presented in \citet{karoly2025jcmtbistrosurveymagnetic}.

In star-forming regions, the alignment efficiency of dust grains can be quantified by the power index of the dependence of the polarization fraction on total intensity, $p\propto I^{-\alpha}$ \citep[][]{2008ApJ...674..304W}.
We investigate the power-law relation between polarization fraction and total intensity using the Ricean Mean Model proposed by \citet{2019ApJ...880...27P} to examine whether dust grains are aligned by magnetic fields or not.

The organization of this paper is as follows. 
In Sect. \ref{sec:obs}, we provide a detailed account of the observed data from JCMT BISTRO and outline the ancillary data used in the estimation of magnetic field strength.
In Sect. \ref{sec:result}, we determine the magnetic field morphology and strength through the analysis of JCMT BISTRO observations.
Using the derived magnetic field strength, we then explore energy contributions from turbulence, gravity, thermal motion, and the magnetic field.
Additionally, we assess the alignment efficiency of dust grains.
Discussions are presented in Sect. \ref{sec:discussion}, and we draw conclusions in Sect. \ref{sec:conclusion}.

\begin{table*}[ht!]
\centering
\caption{Properties of continuum observations}
\begin{tabular}{rccccccc} 
\hline
\hline 
    \multicolumn{1}{b{1.cm}}{} & 
    \multicolumn{2}{b{4cm}}{\centering{Observed Position$^{a}$}} & 
    \multicolumn{1}{b{1.5cm}}{\centering{Duration}} &
    \multicolumn{1}{b{2cm}}{\centering{$\langle \delta I_{450\mu m}\rangle$}$^{b}$} &
    \multicolumn{1}{b{2cm}}{\centering{$\langle \delta I_{850\mu m}\rangle$}$^{b}$} &
    \multicolumn{1}{b{2cm}}{\centering{$\langle \sigma_{QU,450\mu m}\rangle$}$^{b}$} &
    \multicolumn{1}{b{2cm}}{\centering{$\langle \sigma_{QU,850\mu m}\rangle$}$^{b}$} 
\\[-0.1em]
\cmidrule{2-8}\\[-1.7em]
    \multicolumn{1}{b{1.cm}}{\centering Region} & 
    \multicolumn{1}{b{2cm}}{\centering \emph{l} [$\degree$]} & 
    \multicolumn{1}{b{2cm}}{\centering \emph{b} [$\degree$]} & 
    \multicolumn{1}{b{1.5cm}}{\centering{[hr]}} &
    \multicolumn{1}{b{2.0cm}}{\centering{[Jy/beam]}} &
    \multicolumn{1}{b{2.0cm}}{\centering{[Jy/beam]}} &
    \multicolumn{1}{b{2.0cm}}{\centering{[Jy/beam]}} &
    \multicolumn{1}{b{2.0cm}}{\centering{[Jy/beam]}} 
    \\[0.25em]
\hline
    \multicolumn{8}{b{16cm}}{\centering{M20AL018}}
    \\
\hline
    \multicolumn{1}{b{1.cm}}{\centering{Field 1}} &
    \multicolumn{1}{b{2cm}}{\centering 359.883033} & 
    \multicolumn{1}{b{2cm}}{\centering -0.074825} & 
    \multicolumn{1}{b{1.5cm}}{\centering{1.5}} &
    \multicolumn{1}{b{1.5cm}}{\centering{8.01}} &
    \multicolumn{1}{b{1.5cm}}{\centering{0.04}} &
    \multicolumn{1}{b{1.5cm}}{\centering{66.54}} &
    \multicolumn{1}{b{1.5cm}}{\centering{3.99}} 
\\
\hline
    \multicolumn{8}{b{16cm}}{\centering{M17AP074}}
\\
\hline
    \multicolumn{1}{b{1.cm}}{\centering{Field 2}} &
    \multicolumn{1}{b{2cm}}{\centering 0.042601} & 
    \multicolumn{1}{b{2cm}}{\centering -0.010758} & 
    \multicolumn{1}{b{1.5cm}}{\centering{4.0}} &
    \multicolumn{1}{b{1.5cm}}{\centering{7.51}} &
    \multicolumn{1}{b{1.5cm}}{\centering{0.05}} &
    \multicolumn{1}{b{1.5cm}}{\centering{85.92}} &
    \multicolumn{1}{b{1.5cm}}{\centering{5.94}} 
    
\\
    \multicolumn{1}{b{1.cm}}{\centering{Field 3}} &
    \multicolumn{1}{b{2cm}}{\centering 359.944160} & 
    \multicolumn{1}{b{2cm}}{\centering -0.046052} & 
    \multicolumn{1}{b{1.5cm}}{\centering{7.5}} &
    \multicolumn{1}{b{1.5cm}}{\centering{3.51}} &
    \multicolumn{1}{b{1.5cm}}{\centering{0.05}} &
    \multicolumn{1}{b{1.5cm}}{\centering{52.21}} &
    \multicolumn{1}{b{1.5cm}}{\centering{3.89}} 
    
\\
    \multicolumn{1}{b{1.cm}}{\centering{Field 4}} &
    \multicolumn{1}{b{2cm}}{\centering 359.884075} & 
    \multicolumn{1}{b{2cm}}{\centering -0.071629} & 
    \multicolumn{1}{b{1.5cm}}{\centering{4.4}} &
    \multicolumn{1}{b{1.5cm}}{\centering{7.80}} &
    \multicolumn{1}{b{1.5cm}}{\centering{0.04}} &
    \multicolumn{1}{b{1.5cm}}{\centering{59.78}} &
    \multicolumn{1}{b{1.5cm}}{\centering{3.94}} 
\vspace{0.5em}
\\
\hline\hline
\multicolumn{8}{b{16cm}}{Note. The observations from project M20AL018 cover the south region (Field 1); the observations from project M17AP074 cover the north (Field 2), middle (Field 3), and south (Field 4) regions.}\\
\multicolumn{8}{b{16cm}}{$^{a}$ The observed position is expressed in galactic coordinates.}\\
\multicolumn{8}{b{16cm}}{$^{b}$ These values are the root-mean-square (RMS) average of the 4$''$ mosaic maps within the central 3$'$ regions centered on the observed positions.}\\
\end{tabular}
\label{table:OBS_POS}
\end{table*}

\section{Observations} \label{sec:obs}
\subsection{JCMT SCUBA-2/POL-2 Observations} \label{subsec:obs_bistro}

The 450\,$\rm \upmu$m and 850\,$\rm \upmu$m continuum and polarization observations were obtained using the Submillimetre Common-User Bolometer Array 2 (SCUBA-2) equipped with the POL-2 polarimeter on the JCMT 15m telescope (project code: M20AL018, M17AP074).
We observed the 20MC, 50MC, and CND using four pointings (see Table \ref{table:OBS_POS}).
The effective beam sizes of the JCMT are $9\rlap{.}''8$ at 450\,$\rm \upmu$m and $14\rlap{.}''6$ at 850\,$\rm \upmu$m \citep{2013MNRAS.430.2534D}, corresponding to length scales of $\sim0.39$\,pc and $\sim0.59$\,pc, respectively, based on a distance of 8.34\,kpc to Sgr\,A* \citep{2014ApJ...783..130R}.

We reduced the raw data following a three-stage process using \emph{pol2map} routine, which is contained in the SCUBA-2 map-making software \emph{SMURF} \citep{2005ASPC..343...71B, 2013MNRAS.430.2545C}. 
The POL-2 data reduction was described in detail by \citet{2021ApJ...913...85H}.
The final mosaicked maps with a pixel size of $4''$ are created by coadding the Stokes I, Q, U maps.
To achieve better sensitivity, the final debiased polarization catalog is binned to $12''$.
The produced maps of Stokes I, Q, U have units of pW.
To convert the instrumental unit to the physical flux unit of Jy\,beam$^{-1}$, we applied the Flux Conversion Factors (FCFs) of 531 for 450\,$\rm \upmu m$ and 516 for 850\,$\rm \upmu m$, as recommended by \citet{2021AJ....162..191M}.
The FCFs for the observations taken by SCUBA-2 equipped with POL-2 are 1.96 and 1.35 times higher than the ones only taken by SCUBA-2 at 450\,$\rm \upmu m$ and 850\,$\rm \upmu m$, respectively.

We adopt the asymptotic estimator, which introduced by \cite{1974ApJ...194..249W}, to calculate the debiased linear polarization fraction $p$, as follows (\emph{I}, \emph{Q}, and \emph{U} are the Stokes parameters, whose uncertainties are denoted as $\delta I$, $\delta Q$ and $\delta U$, respectively):

\begin{equation}
    p=\frac{PI}{I}=\frac{\sqrt{Q^2+U^2-\sigma_{QU}^2}}{I},
    \label{eqn:p_AS}
\end{equation}
where the weighted mean variance for \emph{Q} and \emph{U}, $\sigma_{QU}^2$, is,
\begin{equation}
    \sigma_{QU}^2=\frac{Q^2\delta Q^2+U^2\delta U^2}{Q^2+U^2}.
    \label{eqn:std_P}
\end{equation}
Here, $PI$ is the debiased polarization intensity.
The $\sigma_{QU}$ indicates the uncertainty of both the debiased and non-biased polarization intensity, $\delta PI$.

If $(Q^2+U^2)<\sigma_{QU}^2$, $p$ will be treated as 0.
The uncertainty on polarization fraction (both debiased and non-debaised) is defined as 
\begin{equation}
    \delta p=\frac{\sqrt{Q^2\delta Q^2+U^2\delta U^2+p^4I^2\delta I^2}}{pI^2},
    \label{eqn:dp_AS}
\end{equation}

In addition, the polarization position angle ($\theta_p$) and its uncertainty ($\delta \theta_p$) are calculated using the following formulae:
\begin{equation}
    \theta_p=\frac{1}{2}\arctan(\frac{U}{Q}),
    \label{eqn:PA}
\end{equation}
\begin{equation}
    \delta \theta_p=\frac{1}{2}\sqrt{\frac{(Q^2\delta U^2+U^2\delta Q^2)}{(Q^2+U^2)^2}}.
    \label{eqn:std_PA}
\end{equation}

\subsection{JCMT CHIMPS2 Observations} \label{subsec:obs_JCMT}

We utilize the $\rm^{12}CO$ (3-2) data from the CO Heterodyne Inner Milky Way Plane Survey 2 (CHIMPS2), obtained with the Heterodyne Array Receiver Program (HARP) on the JCMT \cite[][]{2020MNRAS.498.5936E}, to evaluate the contamination from CO emission. 
These observations cover the CMZ within the range of $-3\degree\le l\le +5\degree$ and $-0\rlap{.}\degree5\le b\le +0\rlap{.}\degree5$.
The velocity range extends from -250\,km\,s$^{-1}$ to 300\,km\,s$^{-1}$ with a resolution of 1\,km\,s$^{-1}$.

\subsection{Contamination of Dust Continuum with CO Emission} \label{subsec:CO_conta}

The $\rm ^{12}CO$ (6-5) line (691.473\,GHz) and $\rm ^{12}CO$ (3-2) line (345.796\,GHz) are close to the centers of the continuum bandpass filters at 450$\,\upmu$m (664\,GHz) and at 850\,$\rm \upmu$m (355\,GHz), respectively, indicating that the CO spectral line emissions may contaminate the continuum observations.
We use the $\rm ^{12}CO$ (3-2) contamination fraction provided by \cite{https://doi.org/10.11570/17.0009} to correct the Stokes I map at 850\,$\upmu$m.
Then, we infer the potential line contamination from $\rm ^{12}CO$ (6-5) to 450\,$\upmu$m continuum observation using $\rm ^{12}CO$ (3-2) contamination fraction under the assumption of local thermodynamic equilibrium \citep[see][]{2012MNRAS.426...23D}.
We assume that the CO emission is unpolarized, which does not affect the Stokes Q and U.


Due to missing flux issue, we did not use BISTRO Stokes I and CHIMPS2 data to estimate the CO contamination.
We have attempted to remove CO contamination during the reduction process using CHIMPS2 $\rm ^{12}CO$ (3-2) observations.
We obtained unreasonable results with negative contamination levels or more than 100\%, however, possibly due to the missing flux problem in POL-2. 
The observing mode used for our BISTRO data is \emph{DAISY}, which is designed for compact sources of $\sim3'$ or smaller. 
The CHIMPS2 observations, which use the \emph{PONG} observing mode, are not suitable for incorporation into our data. 
Detailed descriptions and a discussion of the missing flux issue are provided in Appendix \ref{appendix:CO_removal}.
This issue is also apparent in Sect. \ref{subsubsec:volume_density}, where we use Herschel data to fit the spectral energy distribution (SED).

\subsection{Nobeyama Observations} \label{subsec:c18O}
We apply observations of $\rm C^{18}O$ (1-0) using the 45-m telescope at Nobeyama Radio Observatory (NRO) to analyze the velocity dispersion of the non-thermal component of magnetized turbulence.
These observations have a resolution of 15$''$, similar to the resolutions of our JCMT data.
\citet{2019PASJ...71S..19T} conducted these observations using the FOur-beam REceiver System (FOREST) for the NRO \citep{2016SPIE.9914E..1ZM} during the period of 26-31 January 2016.
The observed region covered the range of approximately $-1\rlap{.}\degree4\le l\le +1\rlap{.}\degree4$ and $-0\rlap{.}\degree35\le b\le +0\rlap{.}\degree35$, using the on-the-fly (OTF) mapping mode. 
The data were smoothed using a 15$''$ Bessel-Gauss function and sampled on a grid of $\rm 7\rlap{.}''5\times 7\rlap{.}''5\times 2\,km\,s^{-1}$.

\subsection{Herschel Hi-GAL} \label{subsec:HiGAL}
We obtained the image of continuum emissions at 160\,$\rm \upmu m$, 250\,$\rm \upmu m$, 350\,$\rm \upmu m$, 500\,$\rm \upmu m$ from the \emph{Herschel} Infrared Galactic Plane Survey \citep[Hi-GAL;][]{2016A&A...591A.149M}, a key program of the \emph{Herschel} satellite \citep[][]{2010A&A...518L...1P}, to determine the dust temperature and opacity spectral index by fitting the SED.
These data are convolved using Gaussian kernels to achieve a common spatial resolution of 36$''$ corresponding to that at 500\,$\rm \upmu m$.

\section{Results} \label{sec:result}
\subsection{Magnetic Field Morphology} 
\label{subsec:B_field_morph}

Polarization at submillimeter wavelengths is thought to originate from asymmetric dust grains aligned with magnetic fields \citep{1951ApJ...114..206D, 1996ApJ...470..551D, 2007ApJ...669L..77L, 2015ARA&A..53..501A}. 
\citet{2019A&A...630A..74M} 
To infer the magnetic field morphology, we rotate the polarization position angles by 90$\degree$ \citep[][]{2015ARA&A..53..501A}.

The question arises of whether we need to be concerned about contamination of our observations by foreground emission.  
JCMT SCUBA-2/POL-2 data are fundamentally insensitive to extended emission due to the necessity of accounting for atmospheric emission, with any astrophysical signal on size scales comparable to or larger than the SCUBA-2 array size being indistinguishable from the atmospheric signal and thus removed in the data reduction process, leading all SCUBA-2 and POL-2 maps to have a zero background \citep[][]{2013MNRAS.430.2545C}. 

The response of SCUBA-2 is thus very strongly weighted towards compact, high-column density structures \citep[e.g.][]{2016MNRAS.463.1008W}, and the small map size and slow scanning speed of POL-2 further suppress the recovery of extended emission \citep[][]{2016SPIE.9914E..03F}.
Our expectation that all of our emission arises from the 20MC, 50MC, and CND, rather than from diffuse extended foregrounds, is supported by the analysis of PILOT observations by \citet{2019A&A...630A..74M}, who found a significant discrepancy between PILOT and SCUPOL (the predecessor to POL-2) observations on the position of the 50MC.  
However, when they performed background subtraction on the PILOT data to mimic the spatial frequencies present in the JCMT SCUPOL data, they found that the two measurements agree well.  
This further suggests that the structures that we observe in the Galactic Center do not have a significant contribution from extended foreground or background emission along the LOS.

\subsubsection{B field Orientations from BISTRO Observations}
\label{subsubsec:BISTRO_B_field}

\begin{figure*}[!ht]
\epsscale{1.15}
\plotone{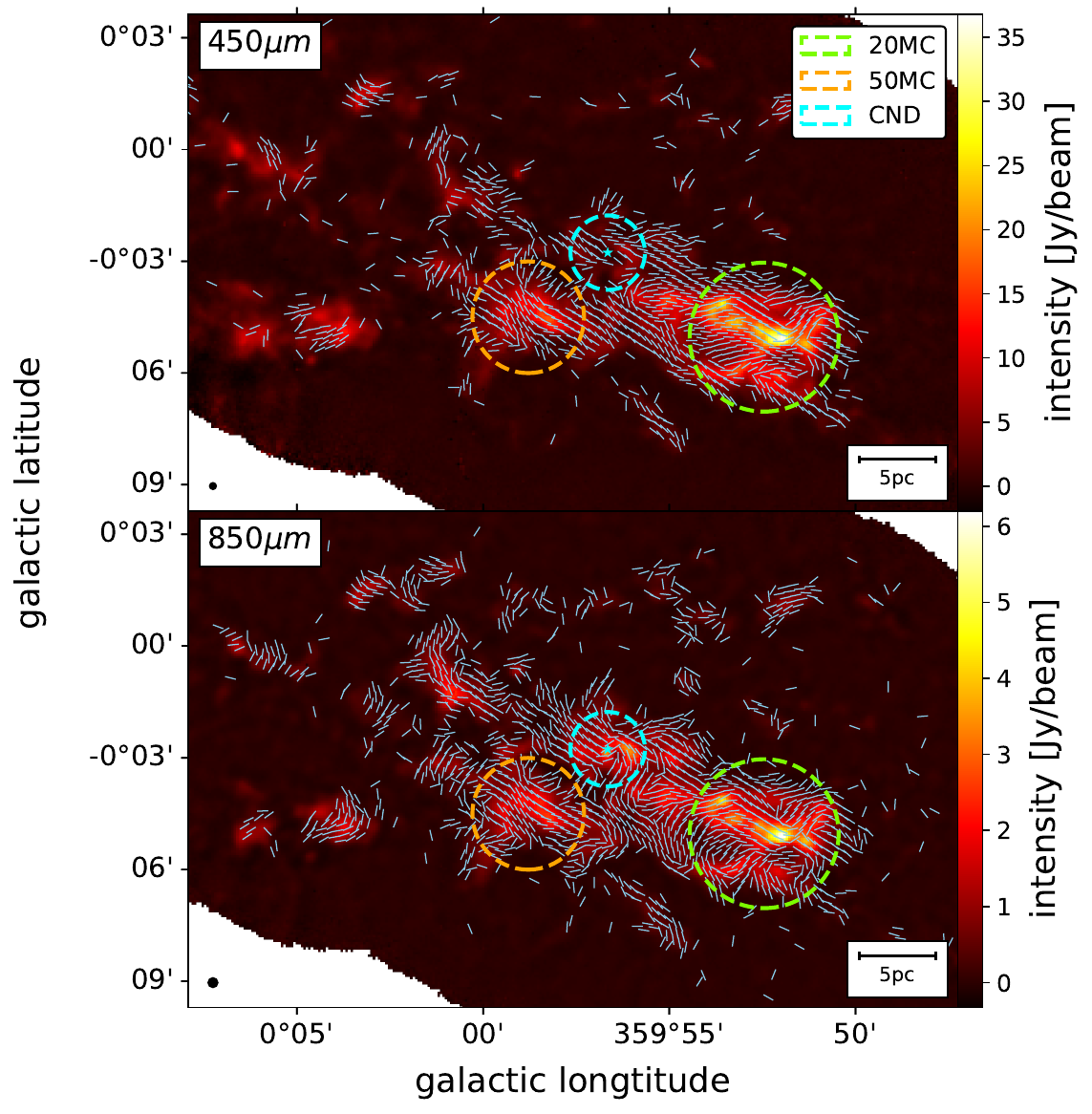}
    \caption{The magnetic field orientations sampled on a 12$''$ grid overlaid on the Stokes I maps containing CO contamination sampled on a 4$''$ grid at 450\,$\rm \upmu m$ and 850\,$\rm \upmu m$.
    White segments represent magnetic field orientations determined by rotating the polarization orientations by 90$\degree$. 
    The polarization half-vectors are selected by the SNR criteria of Stokes I and polarized intensity: $\rm SNR_I > 10$ and $\rm SNR_{PI} > 3$.
    All half-vectors are of uniform length.
    The orange dashed circles mark the position of the 50MC and the green dashed circle marks the position of the 20MC.
    The position of the CND is marked by cyan dashed circle with the central star marking the location of Sgr A*.
    \label{fig:conti_obs}}
\end{figure*}

\begin{figure*} [ht!]
\epsscale{1.}
\plotone{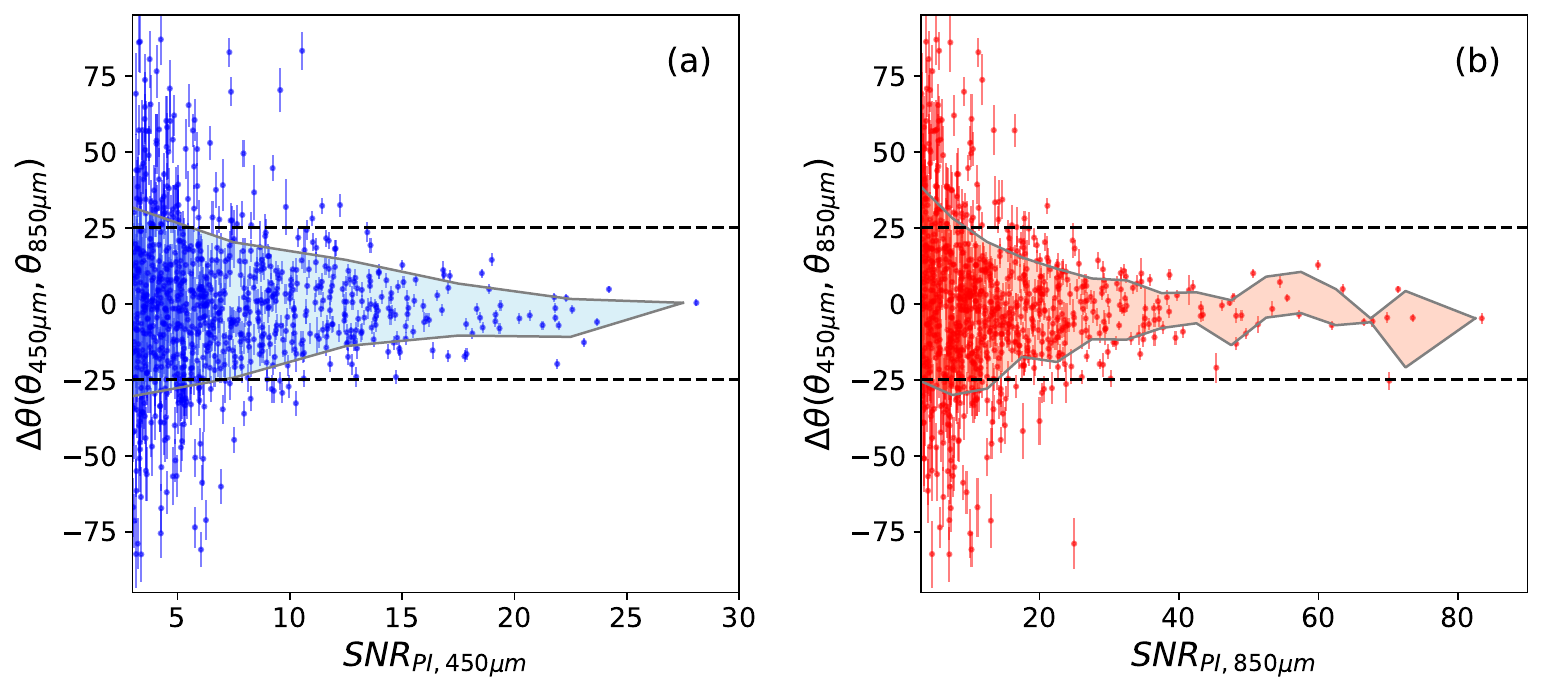}
    \caption{
    Dependence of the magnetic field position angle difference on the SNR of the polarization intensity derived from (a) 450\,$\rm \upmu$m and (b) 850\,$\rm \upmu$m observations. 
    The polarizations are selected by the SNR criteria: $\rm SNR_I > 10$ and $\rm SNR_{PI} > 3$.
    The scatter points represent the position angle differences with corresponding uncertainties for each pixel. 
    The shaded regions in Fig. 3 mark the area between $\overline{\Delta\theta} - \sigma_{\Delta\theta}$ and $\overline{\Delta\theta} + \sigma_{\Delta\theta}$, where $\overline{\Delta\theta}$ and $\sigma_{\Delta\theta}$ indicate the position angle difference and the standard deviation for each 5-unit bin along the axis.
    The dashed lines show the range of $\Delta \theta$ from -25$\degree$ to 25$\degree$.
    \label{fig:PA_diff_vs_PIDPI}}
\end{figure*}

Fig. \ref{fig:conti_obs} illustrates the overall magnetic field morphology traced by POL-2 polarization with a pixel size of 12$''$ at 450\,$\rm \upmu$m and 850\,$\rm \upmu$m, overlaid on the Stokes I map containing CO contamination with a pixel size of 4$''$.
The polarization half-vectors were selected based on the signal-to-noise ratio (SNR) of Stokes I ($\rm SNR_I=I/\delta I$) and polarized intensity ($\rm SNR_{PI}=PI/\delta PI)$.
The criteria include $\rm SNR_I > 10$ and $\rm SNR_{PI} > 3$.
The 20MC, 50MC, and CND are marked by green, orange, and cyan dashed circles, respectively.
Continuum emission closer to the centers of the clouds exhibits lower polarization fractions compared to that observed at the edges.

In Fig. \ref{fig:PA_diff_vs_PIDPI}, we calculate the difference in the magnetic field position angle between the those derived from polarization at 450\,$\rm \upmu$m and 850\,$\rm \upmu$m, ranging from -90$\degree$ to 90$\degree$, and investigate its dependence on the $\rm SNR_{PI}$. 
The angle difference $\Delta \theta$ between two selected position angles, $\theta_{1}$ and $\theta_{2}$, is calculated using the following equation from \citet{2016A&A...586A.135P}:
\begin{equation}
    \Delta \theta(\theta_1, \theta_2)=\frac{1}{2} \arctan{\left(\frac{\sin2\theta_{1}\cos2\theta_{2}-\cos2\theta_{1}\sin2\theta_{2}}{\cos2\theta_{1}\cos2\theta_{2}+\sin2\theta_{1}\sin2\theta_{2}}\right)}.
    \label{eqn:angle_difference}
\end{equation}
Here, $\theta_1$ is the position angle derived from polarization at 450\,$\upmu $m, and $\theta_2$ is the position angle derived from polarization at 850\,$\upmu $m.
We further estimate the average position angle difference, $\overline{\Delta\theta}$, and the standard deviation, $\sigma_{\Delta\theta}$, for each 5-unit bin along the axis.
The shaded regions in Fig. 3 mark the area between $\overline{\Delta\theta} - \sigma_{\Delta\theta}$ and $\overline{\Delta\theta} + \sigma_{\Delta\theta}$.
The shaded areas are mostly located within the $\Delta \theta$ range of -25$\degree$ to 25$\degree$ when the $\rm SNR_{PI}$ is higher than 15 for the 450\,$\upmu$m data and higher than 30 for the 850\,$\upmu$m data, respectively.
This indicates that the magnetic fields inferred from the 450\,$\rm \upmu$m and 850\,$\rm \upmu$m data exhibit greater consistency in regions with high $\rm SNR_{PI}$.
Additionally, low-$\rm SNR_{PI}$, for example where $\rm SNR_{PI, 850\upmu m}$ is below about 30, exhibit more significant position angle differences.

The evident angle difference possible arises from a combination of astrophysical and technical factors, including the beam-size discrepancies and the inherent challenges of 450\,$\upmu\mathrm{m}$ observations.
In general, dense molecular clouds in the CMZ likely have multiple layers of gas and dust, each with a distinct magnetic field orientation, such that the observed polarization angle reflects the vector sum of these layers.
If the LOS contributions at 450\,$\upmu $m differ from those at 850\,$\upmu$m---for instance, if the 450\,$\upmu$m emission selectively samples warmer layers---then the measured polarization angels need not consistent with each other.
Given the nature of CMZ gas, especially toward the CND, some polarization segments at 850\,$\upmu$m could even trace synchrotron emission \citep[see, e.g., Figure 2(b) in ][]{2024PASJ...76..960S}.
Verifying each polarization segment goes beyond the scope of this overview paper, but a paper addressing these topics is currently in preparation by BISTRO members (Ray S. Furuya in preparation).

\begin{figure*} [ht!]
\epsscale{0.95}
\plotone{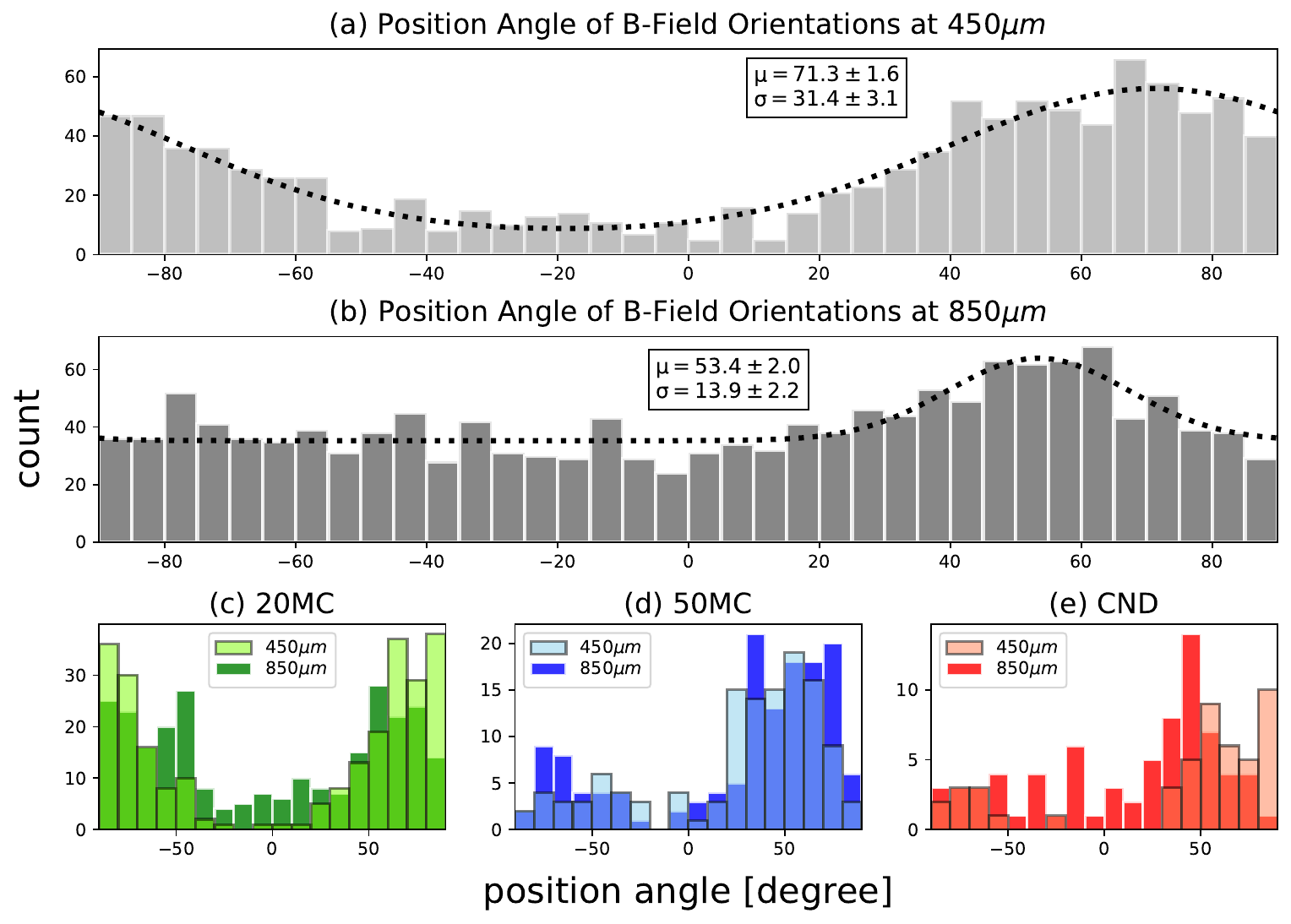}
    \caption{Histogram of the magnetic field orientations derived from (a) 450\,$\rm \upmu m$ and (b) 850\,$\rm \upmu m$ polarization observations. 
    The polarization half-vectors are selected by the criteria $\rm SNR_I > 10$ and $\rm SNR_{PI} > 3$.
    The dark dotted lines show the fitting results from a single Gaussian model.
    Panels (c), (d), and (e) show the histograms of the magnetic fields position angles within the circular regions toward the 20MC, 50MC, and CND, as depicted in Fig \ref{fig:conti_obs}, respectively.
    Magnetic field position angles range from -90$\degree$ to 90$\degree$, with 0$\degree$ pointing north in Galactic coordinates and $\pm90\degree$ aligned with the Galactic plane.}
    \label{fig:Hist_PA}
\end{figure*}

\citet{2024ApJ...969..150P} confirmed a bimodal distribution on the histogram of the CMZ magnetic field orientations using SOFIA/HAWC+ polarization observations at 214\,$\rm \upmu m$.
Their results revealed two distinct magnetic field components: one aligned parallel and the other perpendicular to the Galactic plane.
In Fig. \ref{fig:Hist_PA}, we show histograms of the magnetic field orientations, which are selected by the criteria $\rm SNR_I > 10$ and $\rm SNR_{PI} > 3$, for the entire region at 450\,$\rm \upmu$m and 850\,$\rm \upmu$m in panels (a) and (b), respectively, as well as for each distinct structure in panels (c), (d), and (e).
The position angles for the 20MC, 50MC, and CND are extracted within three circular regions with radii of 1$'$\,($\sim2.4$\,pc), 1.5$'$\,($\sim3.6$\,pc), and 2$'$\,($\sim4.8$\,pc), respectively, as shown in Fig. \ref{fig:conti_obs}.
The magnetic field orientations do not show significant differences in these regions.
Our data, however, do not display a similar bimodal distribution at both 450\,$\rm \upmu$m and 850\,$\rm \upmu$m; instead, they exhibit only one dominant direction.
Therefore, we apply a single-Gaussian model to fit the data, as described by the following equation:
\begin{equation}
    y=H+Ae^{\frac{-(x-\nu)^2}{2\sigma^2}},
    \label{eqn:PA_hist}
\end{equation}
where x is the polarization angle ranging from -90$\degree$ to +90$\degree$, y is the count value, H is the offset above zero, $A$ is the amplitude of the Gaussian peak, $\nu$ is the mean orientations, and $\sigma$ is the standard deviation of the Gaussian peak, which indicates how the position angle distribution spreads.

The best-fit Gaussian peaks are located at position angles of 71.3$\degree$ and 53.4$\degree$ with $\sigma$ of 31.4$\degree$ and 13.9$\degree$ for the 450\,$\rm \upmu$m and 850\,$\rm \upmu$m data, respectively, as shown by the dashed lines in Fig. \ref{fig:Hist_PA} panels (a) and (b).
The fitting result for the 450\,$\rm \upmu$m data indicates that the overall magnetic fields have a position angle difference of $\sim$20$\degree$ relative to the Galactic plane. 
This value is consistent with one of the components in the bimodal distribution proposed by \citet{2024ApJ...969..150P}.
In contrast, the magnetic field derived from 850\,$\rm \upmu$m polarization observations tends to exhibit a larger position angle difference ($\sim$40$\degree$) relative to the Galactic plane.
Additionally, the histogram of the 850\,$\rm \upmu$m data shows a general offset H of $\sim$35 above zero on the y-axis, suggesting a more complex or disordered magnetic field morphology.
In panels (c), (d), and (e), we present histograms of the magnetic field position angles within the circular regions corresponding to the 20MC, 50MC, and CND, all of which are bright dust structures in the inner region of the CMZ, as shown in Fig. \ref{fig:conti_obs}.

\begin{figure*} [ht!]
\epsscale{1.1}
\plotone{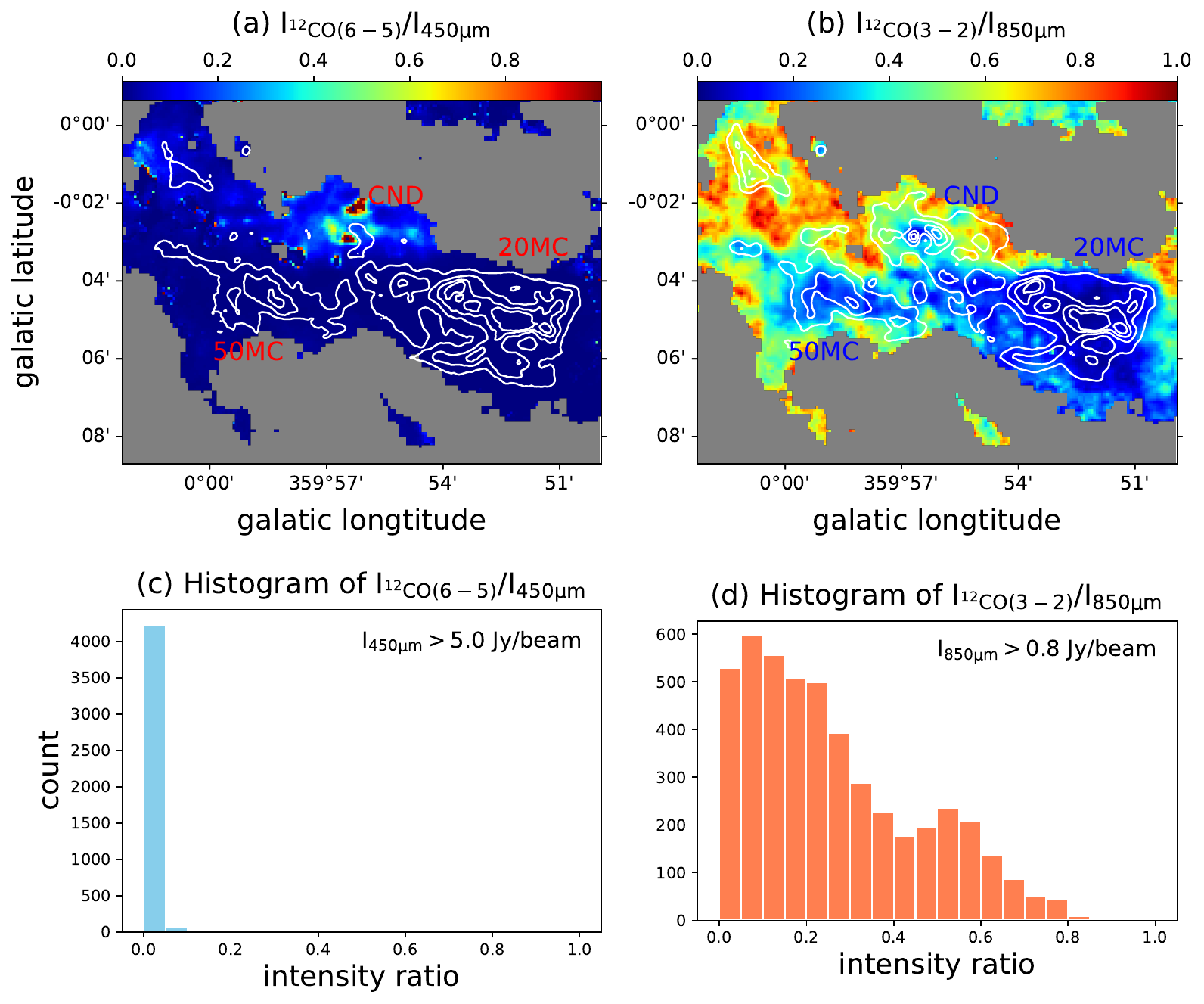}
    \caption{
    The intensity ratio between the CO gas emission and continuum observations, indicating the contamination level by (a) $\rm ^{12}CO$ (6-5) and (b) $\rm ^{12}CO$ (3-2).
    The contours show the 450\,$\rm \upmu m$ intensities of [5, 10, 15, 20]\,$\rm Jy\, beam^{-1}$ in panel (a) and the 850\,$\rm \upmu m$ intensities of [0.8, 1.6, 2.4, 3.2]\,$\rm Jy\, beam^{-1}$ in panel (b).
    Panel (c) illustrates the $\rm ^{12}CO$ (6-5) contamination levels for 450$\rm\ \upmu m$ while $\rm I_{450\upmu m}>$5 Jy/beam.
    Panel (d) shows the $\rm ^{12}CO$ (3-2) contamination levels for 850$\rm\ \upmu m$ while $\rm I_{850\upmu m}>$0.8 Jy/beam.
    \label{fig:CO_Gas_Fraction}}
\end{figure*}

\begin{figure*}[!ht]
    \centering
	\begin{subfigure}{}
	   \includegraphics[width=0.9\linewidth]{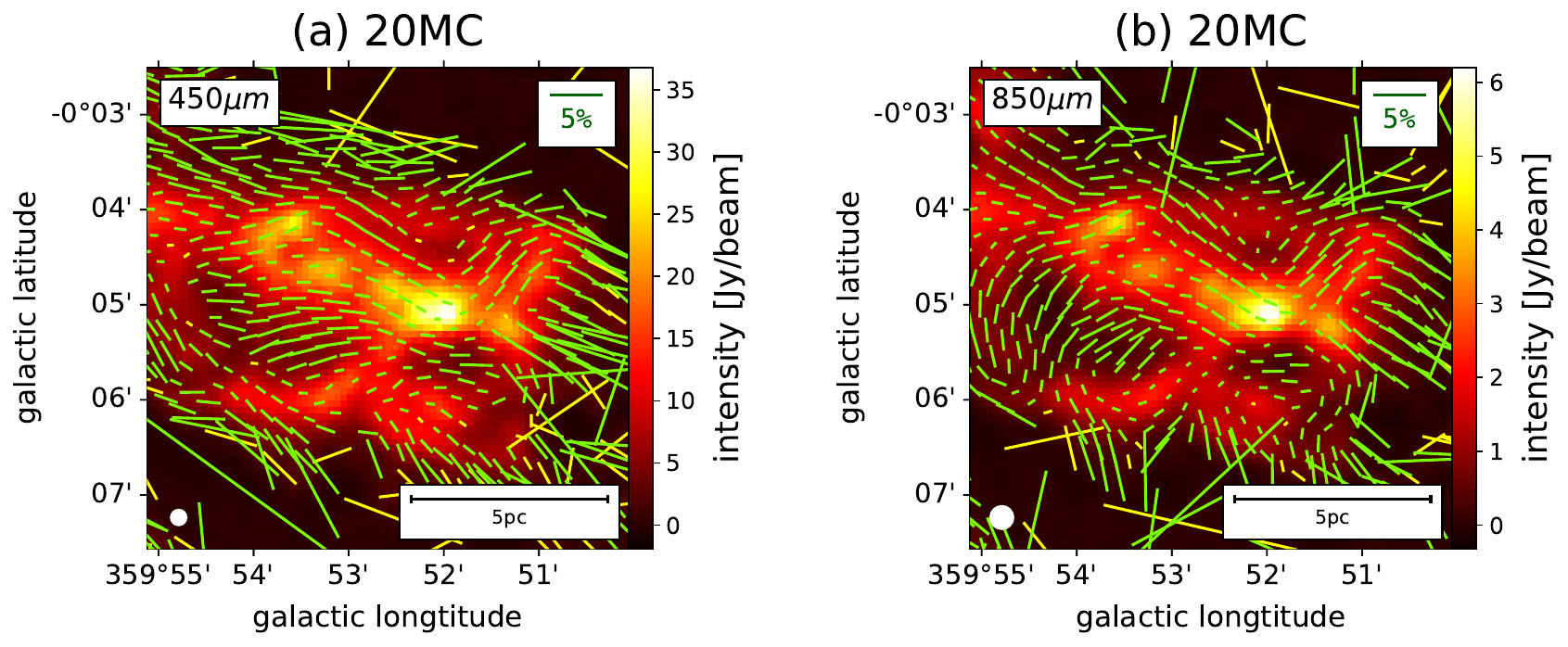}
	\end{subfigure}
	\begin{subfigure}{}
	   \includegraphics[width=0.9\linewidth]{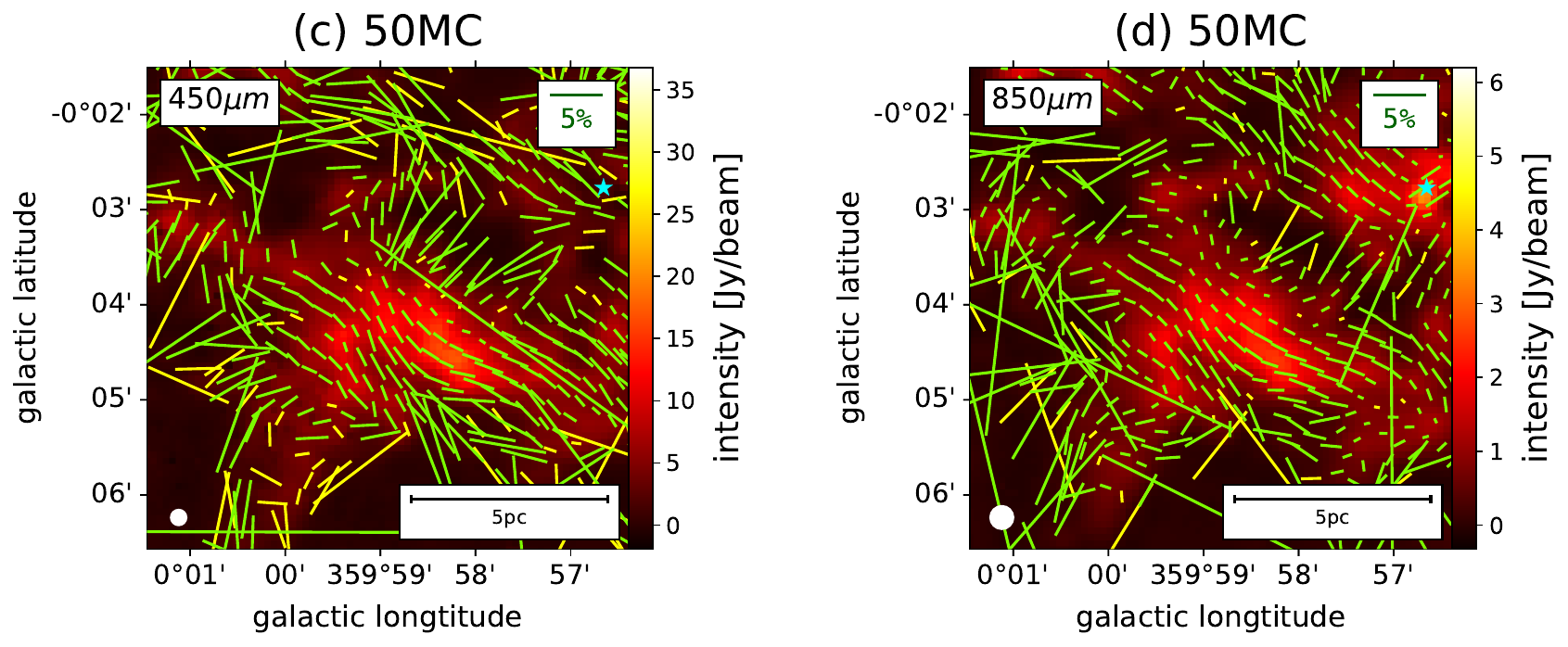}
	\end{subfigure}
	\begin{subfigure}{}
	   \includegraphics[width=0.9\linewidth]{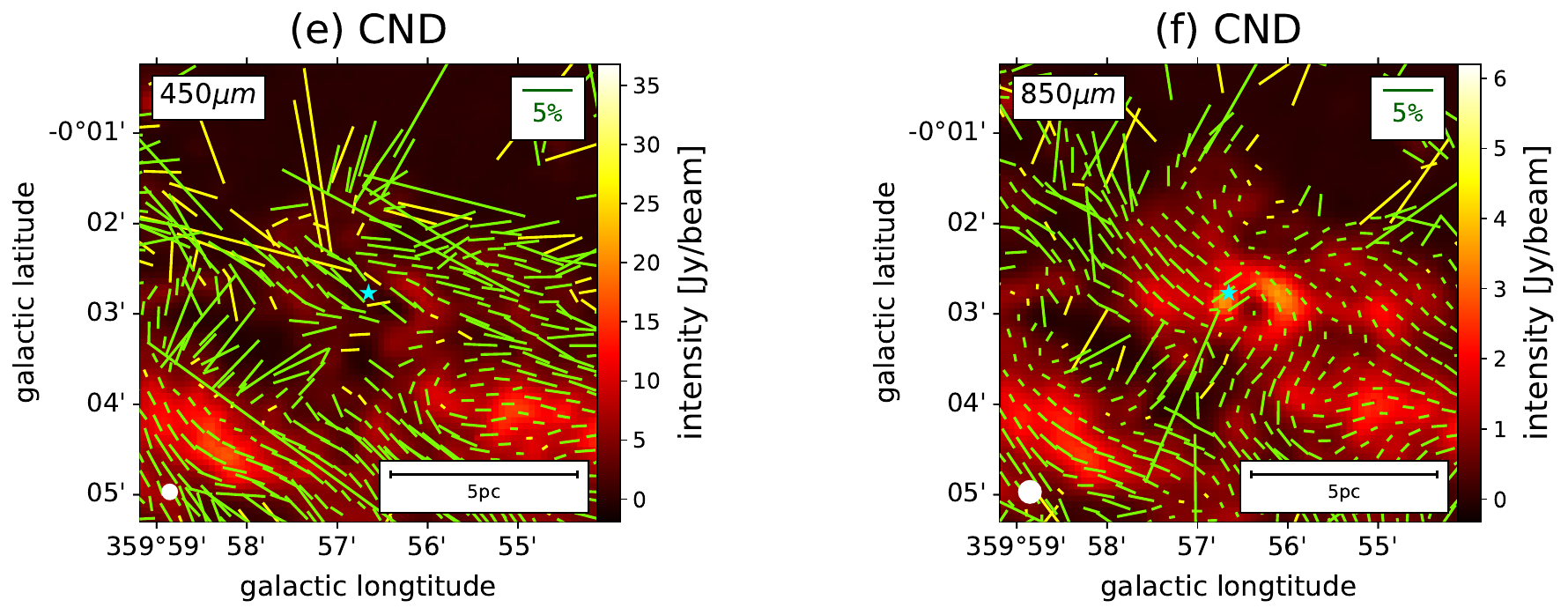}
	\end{subfigure}
	\caption{Zoomed-in maps of three distinct regions: 20MC in panels (a) and (b), 50MC in panels (c) and (d), and CND in panels (e) and (f), at 450\,$\rm \upmu m$ and 850\,$\rm \upmu m$, respectively.
    Magnetic field orientations are represented by line segments, with their lengths proportional to the polarization fraction for a SNR higher than 3 (green) or between 2 and 3 (yellow), respectively. 
    The lengths of the half-vectors are proportional to the polarization fraction obtained using the CO-removed intensity maps.
    }
	\label{fig:zoomin_Bfield}
\end{figure*}

Fig. \ref{fig:CO_Gas_Fraction} shows the contamination level by $\rm ^{12}CO$ (6-5) and $\rm ^{12}CO$ (3-2) proposed by \citet{2018ApJS..234...22P} in panels (a) and (b).
In the vicinity of the CND, dust observations are significantly impacted by CO line emission.
We use these maps to eliminate the impact of CO emission.
Fig. \ref{fig:zoomin_Bfield} gives close-ups of the 20MC, 50MC, and CND to highlight the structure of their magnetic fields.
Magnetic field orientations are represented by orange and yellow line segments with SNR criteria of polarization intensity, $\rm{SNR_{PI}}>3$ and $2\leq \rm{SNR_{PI}}<3$, respectively.
The lengths of the line segments in Fig. \ref{fig:zoomin_Bfield} are proportional to the polarization fractions, which are obtained using the CO-removed intenisty maps.
The position angles reveal a highly ordered magnetic field morphology within each region.
In the 20MC region, the magnetic field orientations align with the Galactic Plane, as indicated by the peak in the histogram in panel (c) of Fig. 4 near a position angle of 90$\degree$.
From panels (a) and (b) in Fig. 5, the field morphology appears to follow the skeleton of the dust structures.
Near the 50MC, the magnetic field points northwest in the northern region and northward in the southern region, with a clear transition at the center.
In the vicinity of the CND, a distinct spiral magnetic field morphology is observed, with its center point roughly consistent with Sgr\,A*.

\begin{figure*} [ht!]
\epsscale{1.1}
\plotone{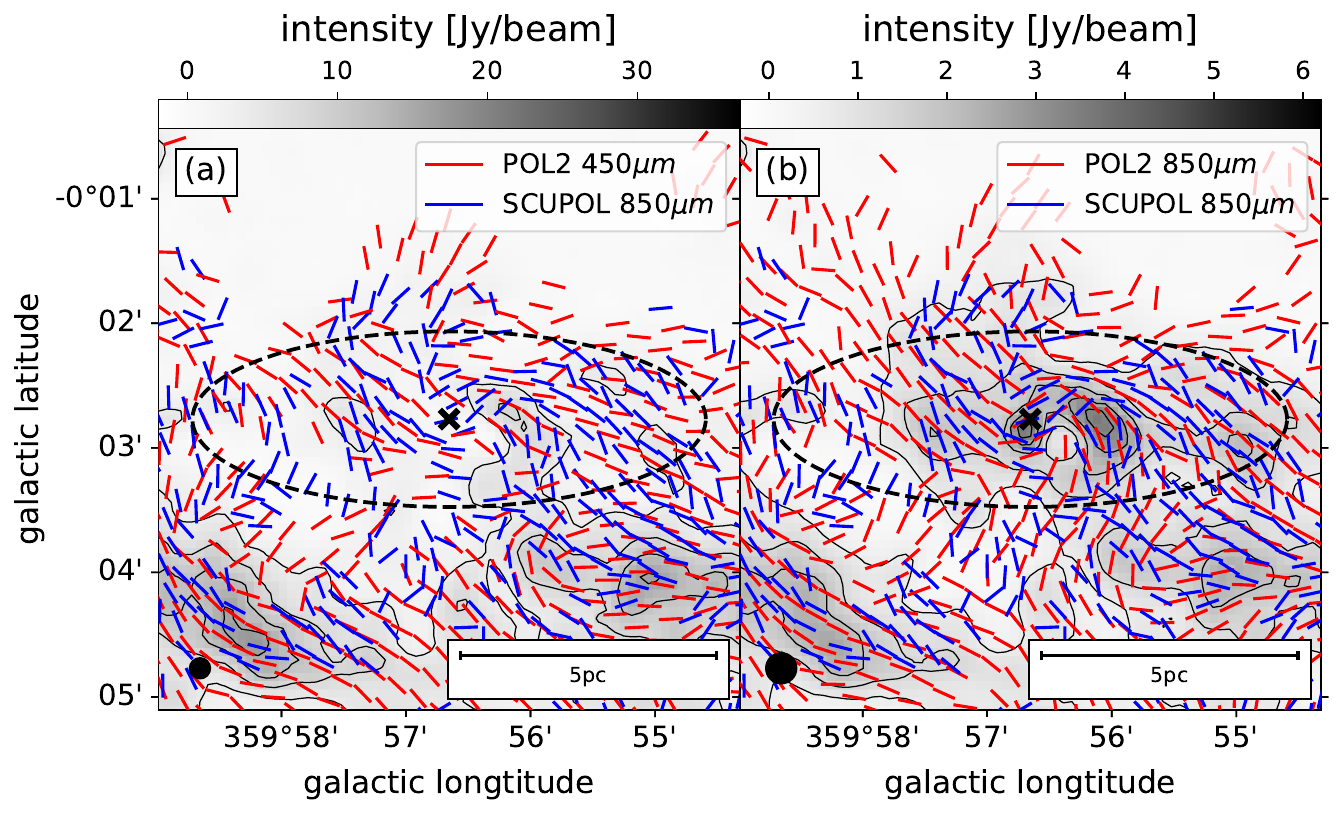}
    \caption{Magnetic field orientations determined by BISTRO 450\,$\rm \upmu m$ and 850\,$\rm \upmu m$ observations (red segments) compared to those determined by SCUPOL 850\,$\rm \upmu m$ observations (blue segments) in the vicinity of the CND.
    All segments are of uniform length.
    Both POL-2 and SCUPOL polarization data are selected by the criteria of $\rm SNR_I > 10$ and $\rm SNR_{PI} > 3$.
    The dashed ellipse, with a 4 pc major axis aligned with the Galactic Plane and an aspect ratio of 0.34, outlines the boundary of the CND.
    \label{fig:CND_B_zoomin}}
\end{figure*}

\subsubsection{Comparison of B Field Position Angles in the CND with SCUPOL Observations}
\label{subsubsec:CND_SCUPOL}

The spiral magnetic field structure (Panel (a) and panel (b) in Fig. \ref{fig:zoomin_Bfield}) had been previously observed at 850\,$\rm \upmu m$ using the polarimeter SCUPOL on the JCMT \citep{2018ApJ...862..150H}.
In Fig. \ref{fig:CND_B_zoomin}, we make a comparison between the magnetic field orientations obtained with POL-2 observations (red segments) and those with SCUPOL observations (blue segments).
We refer the schematic of the CND given by \citet{2019ApJ...885L..20H} to define the boundary.
The CND is thought to be a torus inclined at an angle of $\sim 70\degree$ along the axis parallel to the Galactic Plane.
We define the boundary of the CND as an ellipse with a major axis of 4\,pc and an aspect ratio of 0.34.
The ellipse is centered on Sgr\,A* and aligned with the Galactic Plane.
Between the POL-2 and SCUPOL observations, the magnetic field position angles are a similar in the elliptical region, except for the southeastern part.

\begin{figure*} [ht!]
\epsscale{1.1}
\plotone{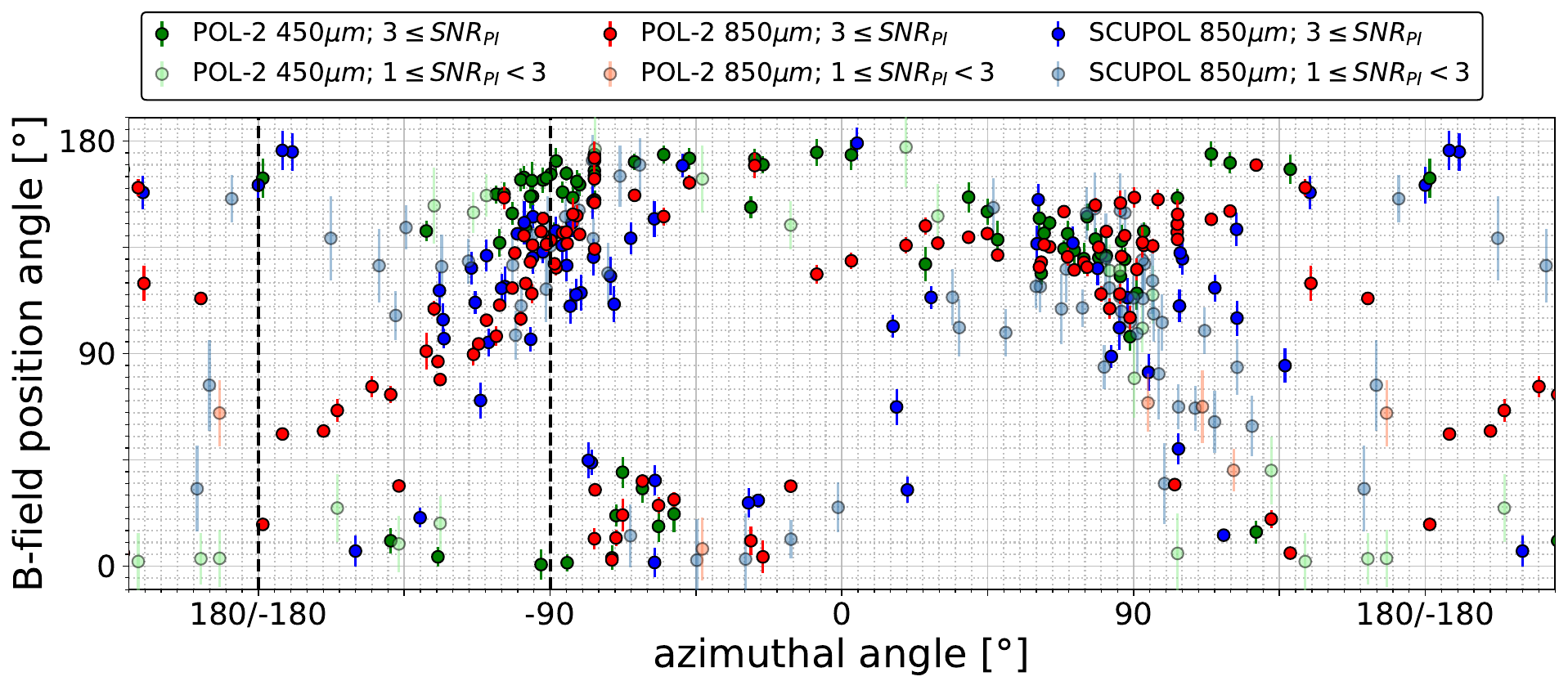}
    \caption{The magnetic field position angle versus azimuthal angle in the CND.
    Azimuthal angle ranges from -180$\degree$ to 180$\degree$ with the north direction in galactic coordinate defined as 0$\degree$.
    Position angles are displayed in the range from 0$\degree$ to 180$\degree$.
    The data enclosed between the two dashed lines represent observations from the southwest region related to Sgr\,A*.
    The different colors of the data points represent various observations and separate the data into polarization intensity SNR ranges of $1 \leq \rm{SNR_{PI}} < 3$ and $\rm{SNR_{PI}}\ge3$.
    \label{fig:PA_vs_AZ_CND}}
\end{figure*}

We investigated the variation in magnetic field position angles with azimuthal angle in the CND, as shown in Fig. \ref{fig:PA_vs_AZ_CND}.
To illustrate the difference between the position angles observed by POL-2 and SCUPOL, we present the data with different SNR of the polarization intensity using different colors. 
The azimuthal angle ranges from -180$\degree$ to 180$\degree$, increasing counterclockwise, with the north direction in the galactic coordinates defined as 0$\degree$.
To analyze the trend of position angle variation within the azimuthal angle range of -180° to -90$\degree$, where the position angle increases from $\sim$0$\degree$ as the azimuthal angle increases, we show the position angle in the range of 0$\degree$ to 180$\degree$.
In the southeastern region (between $-90\degree$ and $-180\degree$), the position angles from the POL-2 and SCUPOL observations are not well-matched. 
The $\rm SNR_{PI}$ of POL-2 850\,$\rm \upmu$m observations are generally higher than 3 in this region. 
The referring magnetic field position angles rotate counterclockwise as the azimuthal angle increases from $-180\degree$ to $-90\degree$, corresponding to the red points rising between the dashed lines.
SCUPOL observations with $\rm SNR_{PI} > 3$ are in good agreement with our data.
Data points with $\rm 1 \le SNR_{PI} < 3$, however, are aligned with different orientations.

Our higher quality observations consequently reveal a more pronounced left-handed spiral morphology in the CND.

\begin{figure*} [ht!]
\epsscale{1.1}
\plotone{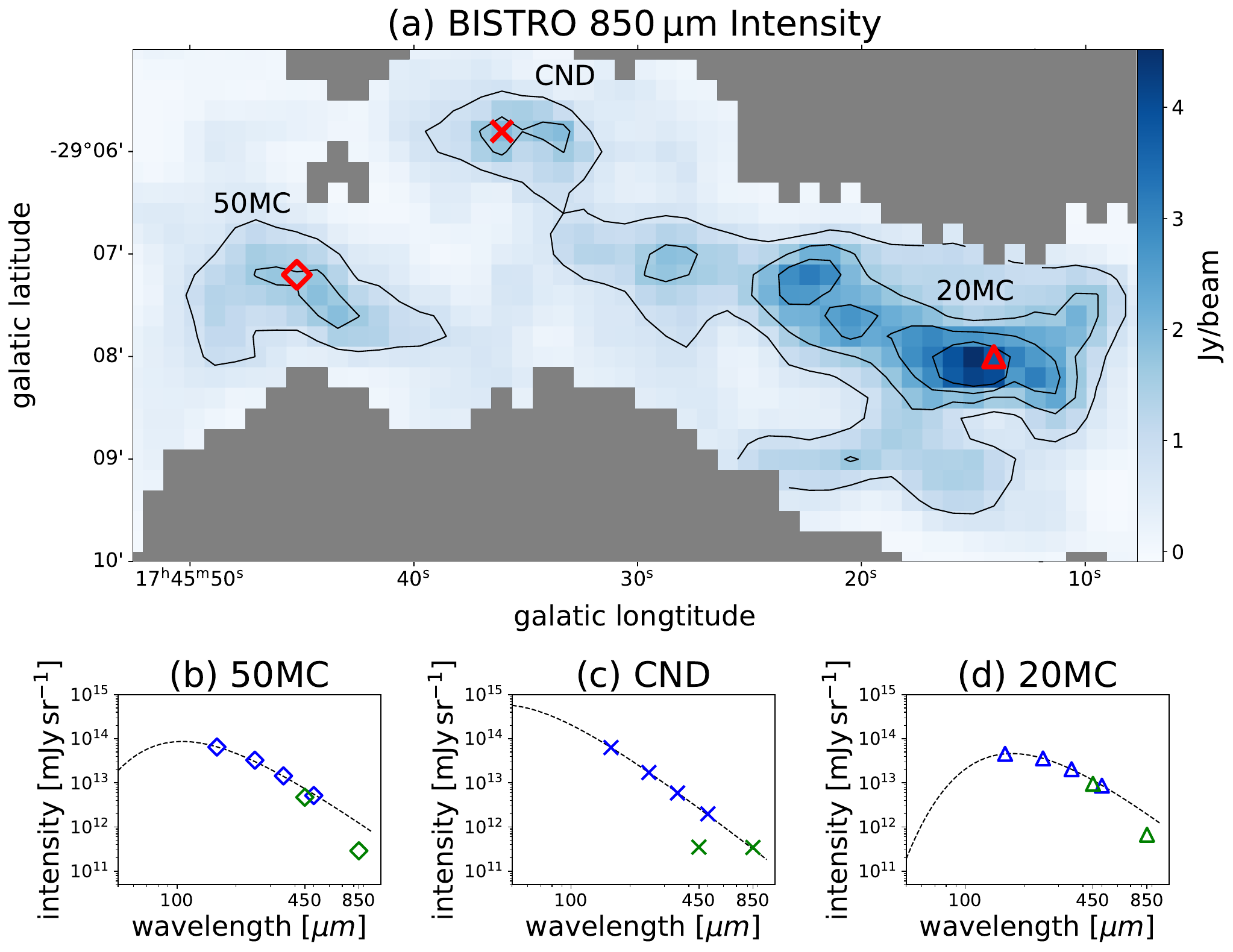}
    \caption{Three cases of the SED fitting toward the 50MC, CND, and 20MC, shown in panels (b), (c), and (d). 
    The panel (a) shows the CO-removed intensity at 850\,$\rm \upmu m$ with contours corresponding [0.8, 1.6, 2.4, 3.2]\,$\rm Jy\, beam^{-1}$.
    Positions in the vicinity of 20MC, 50MC, and CND are selected as examples and shown in right panel.
    The right panel displays the SED fitting results (black dashed line) using HiGAL observations, with HiGAL data represented in blue and  CO-removed BISTRO intensity in green. 
    In panels (b), (c), and (d), all uncertainties are less than 1\% and are therefore not shown.
    \label{fig:SED_3fittingline}}
\end{figure*}

\subsection{Magnetic Field Strength} \label{subsec:B_field_stren}

We use the Davis-Chandrasekhar-Fermi (DCF) method \citep[][]{1951PhRv...81..890D,1953ApJ...118..113C} to estimate the POS magnetic field strength by assuming the equipartition between the kinetic energy of turbulence and the fluctuating magnetic energy.
The POS uniform magnetic field strength is estimated with:
\begin{equation}
    B_{POS}=Q'\sqrt{4\pi \rho}\ \frac{\sigma_\nu}{\sigma_\theta}
    \label{eqn:DCF_B_Pattle}
\end{equation}
Here, $\rho$ is the gas density ($\rho=\mu_{H_2} m_{H} n(H_2)$, where $n(H_2)$ is the number density of molecular hydrogen, $\rm \upmu_{H_2}$ is the mean molecular weight per hydrogen molecule, and $m_H$ is the mass of hydrogen. We adopt $\rm \upmu_{H_2}=2.86$ from \citet{2013MNRAS.432.1424K}.); $\sigma_{\nu}$ is the non-thermal velocity dispersion in the gas; $\sigma_{\theta}$ is the angular dispersion of the polarization position angles
; and $Q'$ represents a dimensionless factor of order unity, which corrects for the bias in magnetic field strength on scales smaller than the beam. 
According to \citet{2022ApJ...925...30L}, we adopt $Q'\approx 0.28$ with 50\% uncertainty.


\citet{2004mim..proc..123C} also suggested a statistical correction to approximate the total magnetic strength in the three-dimensional space
\begin{equation}
    |B_{tot}|\approx \frac{4}{\pi}B_{POS}.
    \label{eqn:DCF_B_Crutcher}
\end{equation}
Therefore, calculating the magnetic field strength requires values for number density, angular dispersion, and velocity dispersion.

\subsubsection{Volume Density} \label{subsubsec:volume_density}

\begin{figure*} [ht!]
\epsscale{1.1}
\plotone{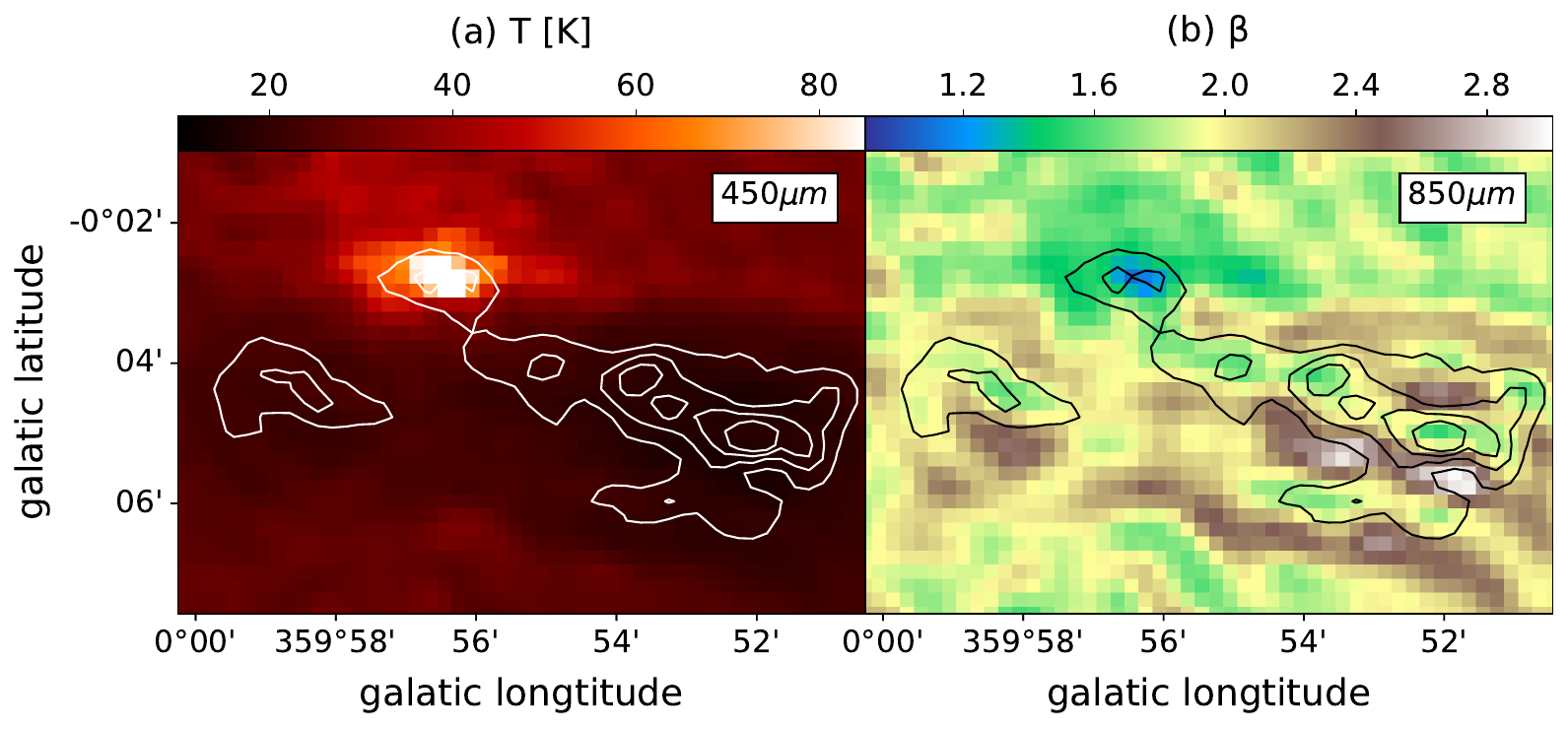}
    \caption{The temperature, opacity spectral index, and column density maps determined by SED fitting using HiGAL observations at wavelength of 160, 250, 350, 500\,$\rm \upmu$m. The contours show the CO-removed intensities at 850\,$\rm \upmu m$ with the values of [0.8, 1.6, 2.4, 3.2]\,$\rm Jy\, beam^{-1}$. 
    The black cross is the position of the Sgr\,A*.
    \label{fig:beta_T}}
\end{figure*}

\begin{figure*} [ht!]
\epsscale{1.1}
\plotone{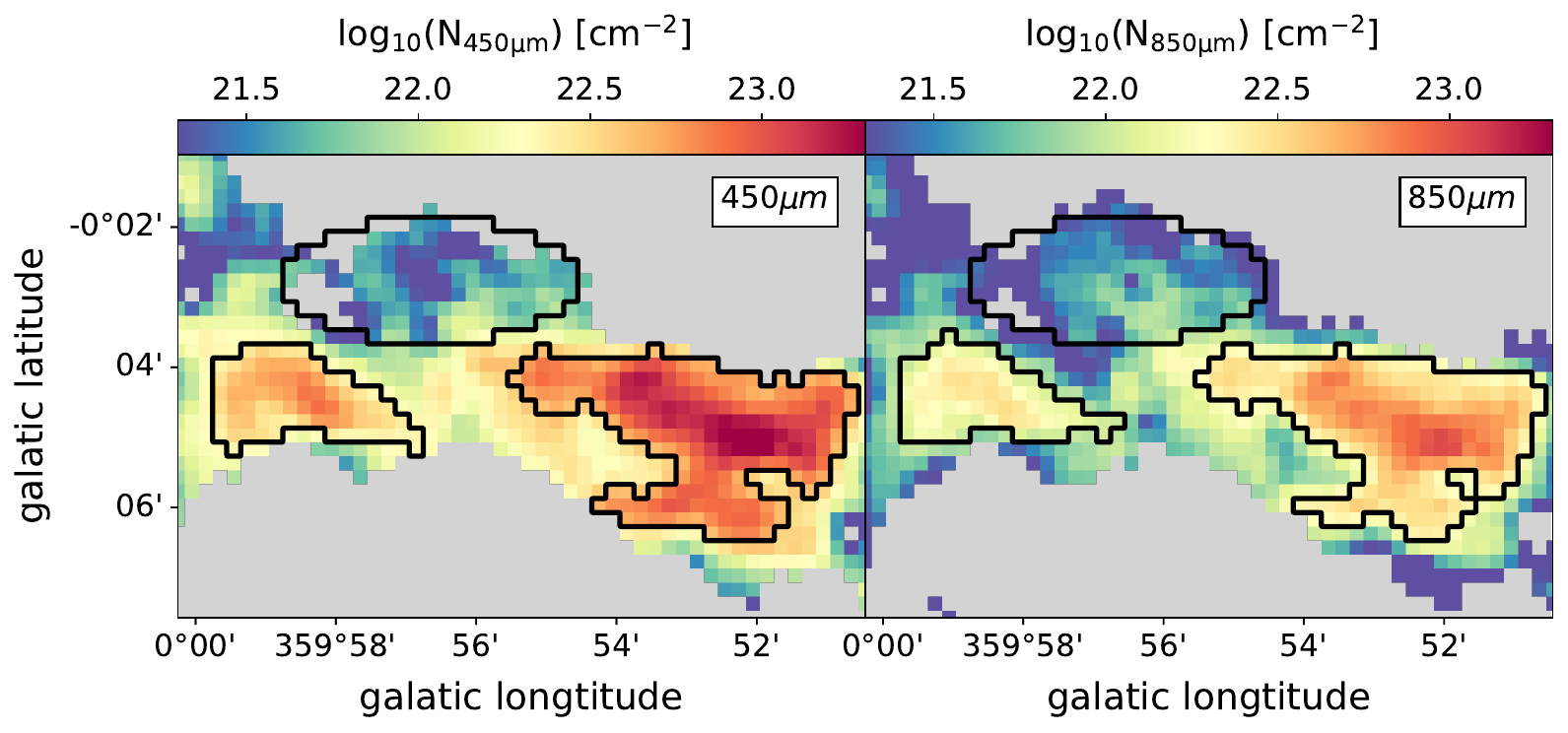}
    \caption{Column density maps estimated by the BISTRO observations at 450\,$\rm \upmu m$ and 850\,$\rm \upmu m$ using the temperature and opacity spectral index parameters shown in Fig. \ref{fig:beta_T}. 
    Black contours depict the boundaries of the 20MC, 50MC, and CND.
    \label{fig:column_density}}
\end{figure*}

\begin{figure*} [ht!]
\epsscale{1.15}
\plotone{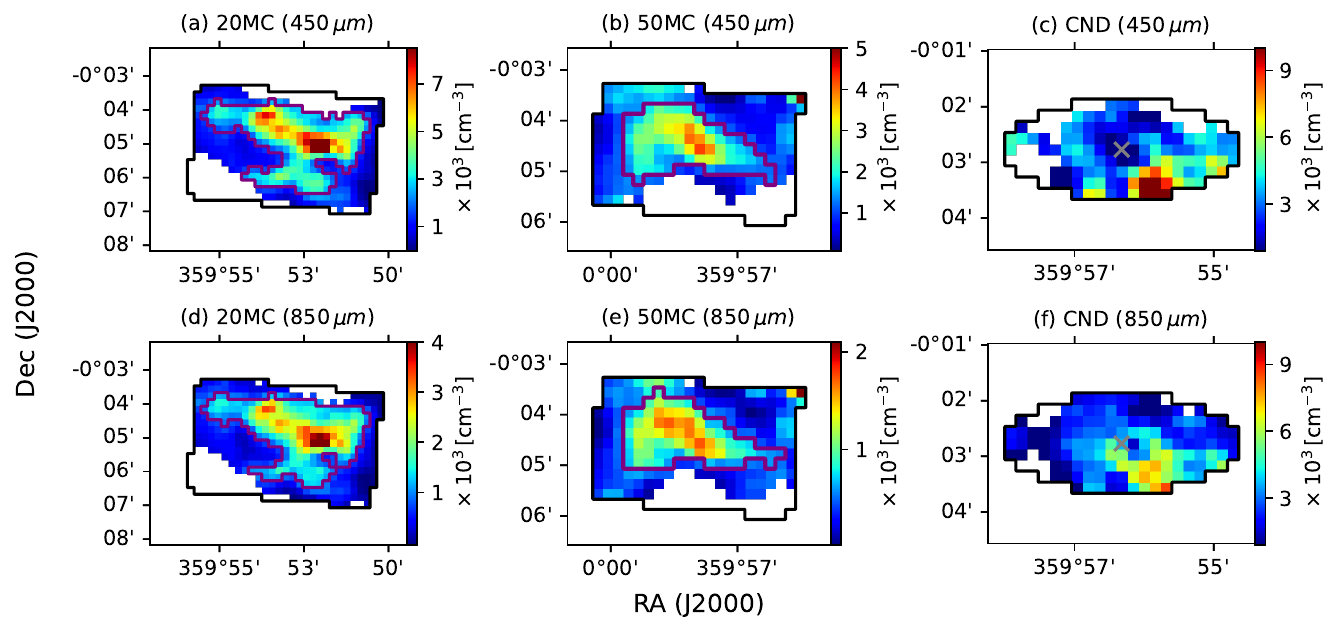}
    \caption{The number density maps in different regions at 450\,$\rm \upmu m$ (a to c) and 850\,$\rm \upmu m$ (d to f), overlaid with the corresponding column density maps.
    Purple solid lines show the structures of 20MC and 50MC determined by \emph{astrodendro}.
    Black solid lines shows the boundaries of the cylinder and ellipse models, used to determine the LOS spatial depth.
    \label{fig:number_density}}
\end{figure*}

As the first step in determining the volume density, we calculate the column density using the CO-removed intensities at 450\,$\rm \upmu m$ and 850\,$\rm \upmu m$.
Under the assumption that the dust emission is optically thin and follows a graybody distribution in the submillimeter observations, the column density ($N_{H_2}$) can be estimated by the opacity spectral index ($\beta$), temperature ($T$), and observed flux density ($S_{\nu}^{beam}$), as follows:
\begin{equation}
    N_{H_2} = \frac{S_\nu^{beam}}{\Omega_A\mu_{H_2}m_H\kappa_\nu B_\nu(T)},
    \label{eqn:RTtransfer}
\end{equation}
where 
\begin{equation}
    B_{\nu}(T)=\frac{2h\nu^3}{c^2}\frac{1}{exp(h\nu/kT)-1},
    \label{eqn:balckbody}
\end{equation}
and 
\begin{equation}
    \kappa_{\lambda}=\kappa_{0}(\frac{1000\,\mu m}{\lambda})^{-\beta}.
    \label{eqn:emissivity}
\end{equation}
Here, $\Omega_A$ is the solid angle of the beam. 
We adopt the dust opacity at the wavelength of 1000\,$\rm \upmu m$, $\kappa_{0}=0.0137\,cm^2\,g^{-1}$ \citep{1994A&A...291..943O}, as the reference in Eq. \eqref{eqn:emissivity}, assuming the gas-to-dust mass ratio of 100.

We can not estimate the column density based solely on the BISTRO observations at two wavelengths because there are three unknowns, including $N_{H_2}$, $\beta$, and $T$.
Therefore, we utilize the continuum data at wavelengths of 160\,$\rm \upmu m$, 250\,$\rm \upmu m$, 350\,$\rm \upmu m$, and 500\,$\rm \upmu m$ from the Hi-GAL survey, which used the \textit{Herschel} Space Observatory to scan the entire Galactic Plane \citep{2017MNRAS.471.2730M}, to fit the opacity spectral index and temperature using Eq. \eqref{eqn:RTtransfer}.
Since the BISTRO continuum intensities at 450\,$\rm \upmu m$ and 850\,$\rm \upmu m$ are not consistent with the SED determined by the HiGAL data, we did not use them simultaneously for SED fitting.
In Fig. \ref{fig:SED_3fittingline}, we convolve all intensity maps with the beam size of Hi-GAL data at 500\,$\rm \upmu m$ (36$''$) and display the intensity profiles with five different wavelengths from both of Hi-GAL (blue) and BISTRO observations (green) at three distinct positions in the vicinity of the 20MC, 50MC, and CND.
The black dashed lines denote the fitting results using solely the Hi-GAL data by the Eq. \eqref{eqn:RTtransfer}, revealing obvious differences with the BISTRO observations.
The lack of consistency between the BISTRO and Hi-GAL data on the SED is caused by SCUBA-2 measurements missing flux from large-scale structures.

Fig. \ref{fig:beta_T} displays the fitting results for the opacity spectral index $\beta$, the temperature, and the column density. 
The $\beta$ value ranges from approximately 1.0 to 2.8, with the lowest value observed around the CND.
The temperature is higher closer to the Galactic Center and decreases near the 20MC and 50MC.
Using this temperature map, the optical depth is $\sim$\,0.1 at 450\,$\rm \upmu m$ and $\sim$\,0.01 at 850\,$\rm \upmu m$, which is consistent with the assumption that the dust emission is optically thin.
We apply the fitted $\beta$ and $T$ to estimate the column density using the BISTRO observations at both 450$\rm \upmu m$ and 850$\rm \upmu m$ wavelengths, shown in Fig. \ref{fig:column_density}.
Near the 20MC and 50MC, the column density exceeds $10^{22}\,cm^{-2}$, and the distributions have similar structures between the wavelengths of 450\,$\rm \upmu m$ and 850\,$\rm \upmu m$.
In the vicinity of the CND, the column density maps show distinct ring-like structures at both 450\,$\rm \upmu m$ and 850\,$\rm \upmu m$, which are consistent with the scenario proposed by previous studies \citep[e.g.][]{2019ApJ...885L..20H}.

To estimate the average physical properties of the 20MC, 50MC, and CND in the following sections (e.g. $n(H_2)$, $\sigma_{\theta}$), we need to define the boundary of each structure.
We use the Python package ${\sl astrodendro}$ (http://www.dendrograms.org/) to compute dendrograms and define the boundaries of the 20MC and 50MC based on the column density maps at both wavelengths.
A lower column density limit of $10^{22}$\,$\rm cm^{-2}$ is applied for the 450\,$\rm \upmu m$ data and $5\times10^{21}$\,$\rm cm^{-2}$ for the 850\,$\rm \upmu m$ data.
Due to the obvious difference in the CND boundaries determined by ${\sl astrodendro}$ at 450\,$\rm \upmu m$ and 850\,$\rm \upmu m$, we adopt the elliptical boundary defined in Sect. \ref{subsubsec:CND_SCUPOL}.
The black contours in Fig. \ref{fig:column_density} depict the  boundaries of the 20MC, 50MC, and CND.

To calculate $n(H_2)$, the LOS spatial depth is necessary.
\citet{2011ApJ...735L..33M} suggested that there is a 100\,pc elliptical and twisted ring around the galactic center, enclosing the 20MC and 50MC.
We model the material distribution around these two massive molecular clouds as cylinders lying on the POS and manually adjust the cylinder radii to encompass the structures determined by $astrodendro$.
The radii and lengths of the cylinders are 4.4\,pc and 15.0\,pc for the 20MC, and 3.0\,pc and 9.6\,pc for the 50MC, respectively.
Using these models, we calculate the LOS distance pixel by pixel and estimate the number density around the 20MC and 50MC.
For the CND region, with the elliptical boundary defined in Sect. \ref{subsubsec:CND_SCUPOL}, we use the typical thickness of $\rm \sim0.4\,pc$ as suggested by \citet{1999ApJ...511..761L} to calculate the number density.
These LOS distances are considered the 50\% errors.
The number density maps in the vicinity of the 20MC, 50MC, and CND are shown in Fig. \ref{fig:number_density}.
Table \ref{table:DCF_parameters} lists the average number densities for these three structures at 450\,$\rm \upmu m$ and 850\,$\rm \upmu m$.
The uncertainty in the number density is calculated by propagating the uncertainties from the observational data, temperature, and opacity spectral index. 
The latter two are based on the 1$\sigma$ errors from the SED fitting.

\subsubsection{Angular Dispersion} \label{subsubsec:angular_dispersion}

\begin{figure*}[!ht]
    \centering
	\begin{subfigure}{}
	   \includegraphics[width=0.9\linewidth]{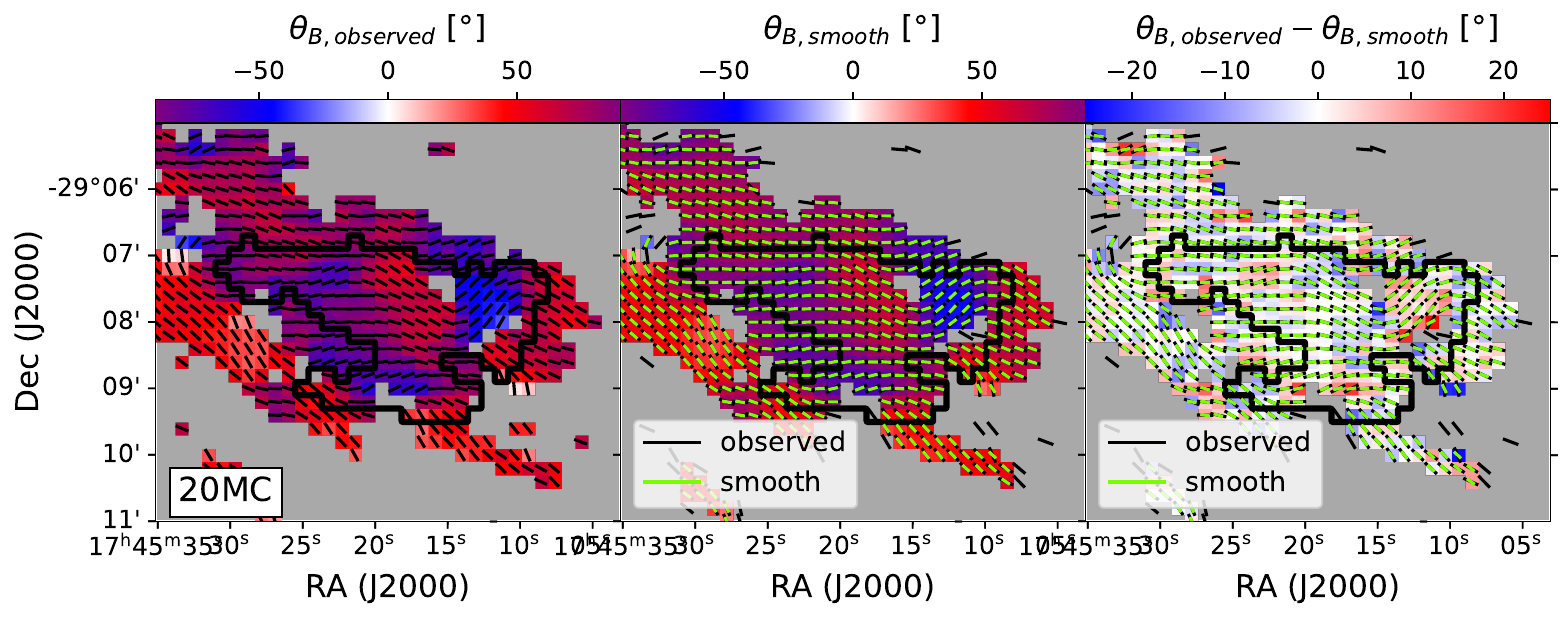}
	\end{subfigure}
	\begin{subfigure}{}
	   \includegraphics[width=0.9\linewidth]{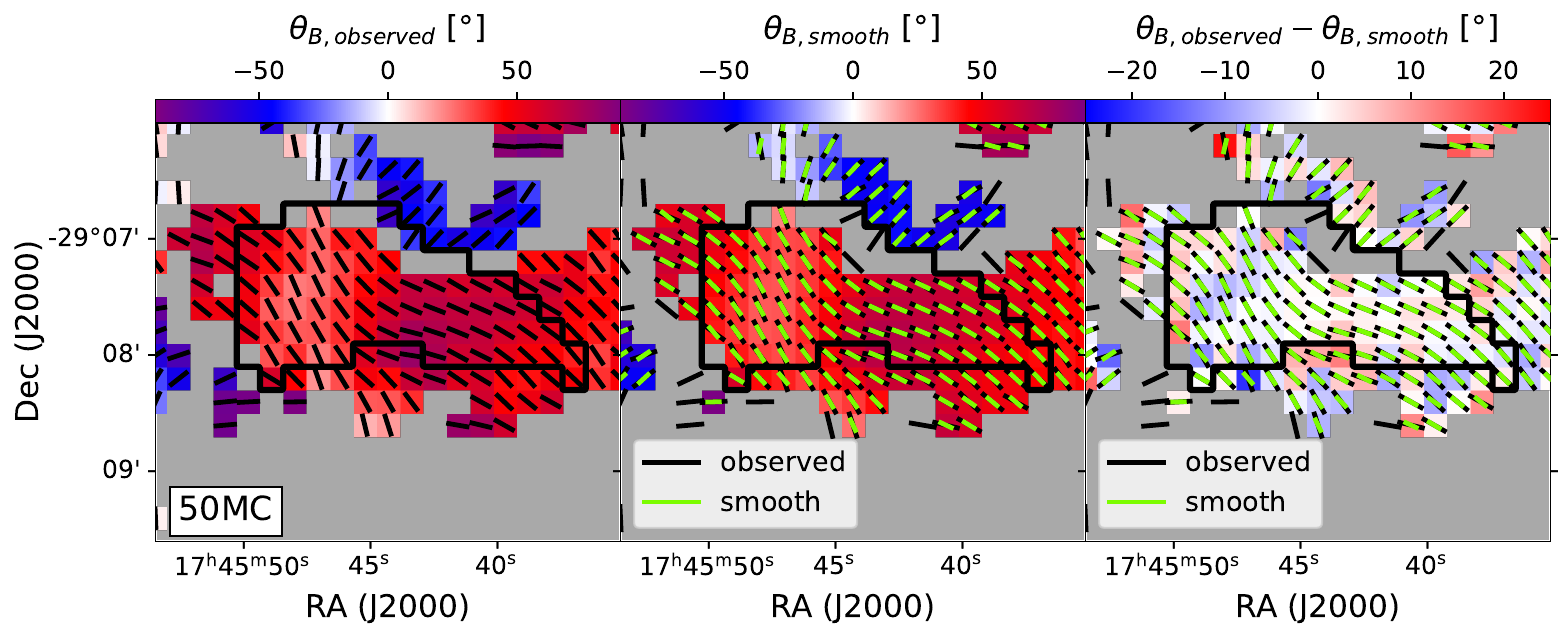}
	\end{subfigure}
	\begin{subfigure}{}
	   \includegraphics[width=0.9\linewidth]{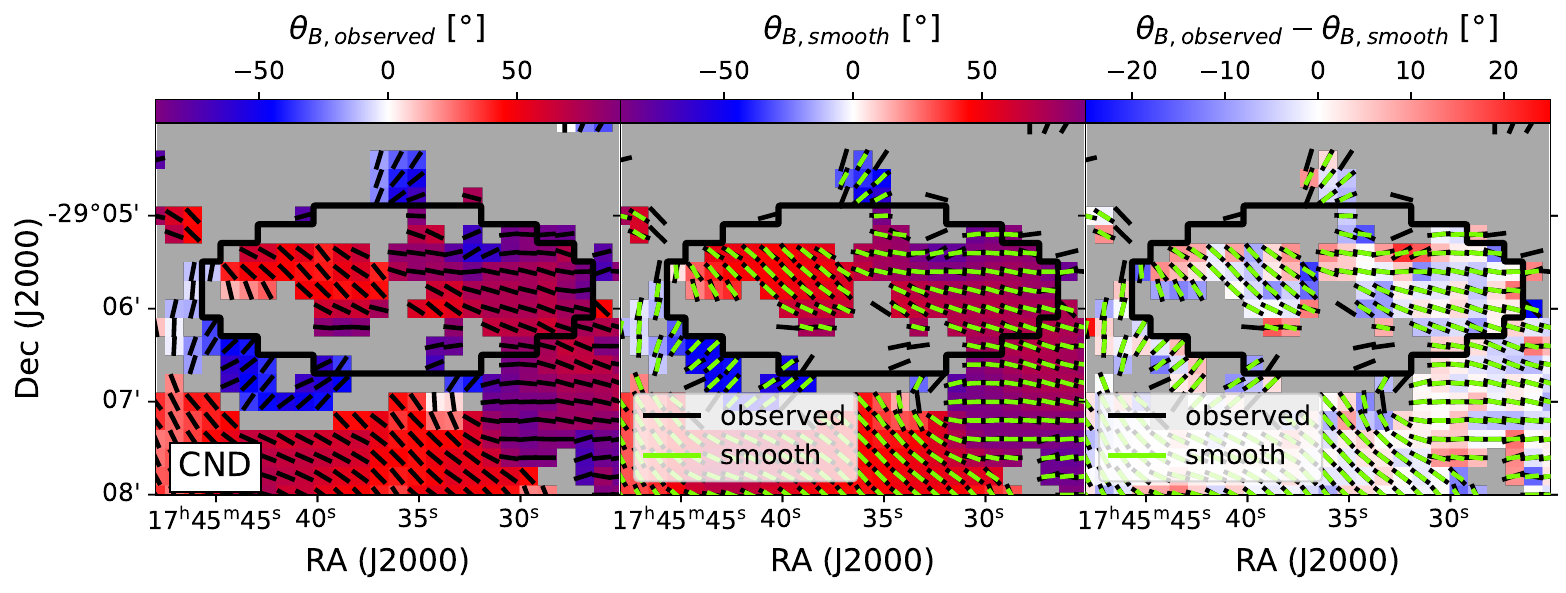}
	\end{subfigure}
	\caption{Position angles of magnetic fields $\theta_B$ determined by 450\,$\rm \upmu m$ polarization in different regions: the 20MC (first row), 50MC (second row), and CND (third row).
    The position angle ranges from -90$\degree$ to 90$\degree$, with 0$\degree$ pointing upward.
    The left panel shows the observed magnetic field position angles $\theta_{B,observed}$; the central panel shows the mean directions of magnetic field orientations $\theta_{B,smooth}$; the right panel shows the angle differences by subtracting the smooth position angles from the observed ones using Eq. \ref{eqn:angle_difference}, $\theta_{B,observed}-\theta_{B,smooth}$. 
    The black segments are the half-vector of $\theta_{B,observed}$, and the green ones are the half-vector of $\theta_{B,smooth}$. 
    The black contours show the boundaries of different structures.}
	\label{fig:smoothPA_450}
\end{figure*}

\begin{figure*}[!ht]
    \centering
	\begin{subfigure}{}
	   \includegraphics[width=0.9\linewidth]{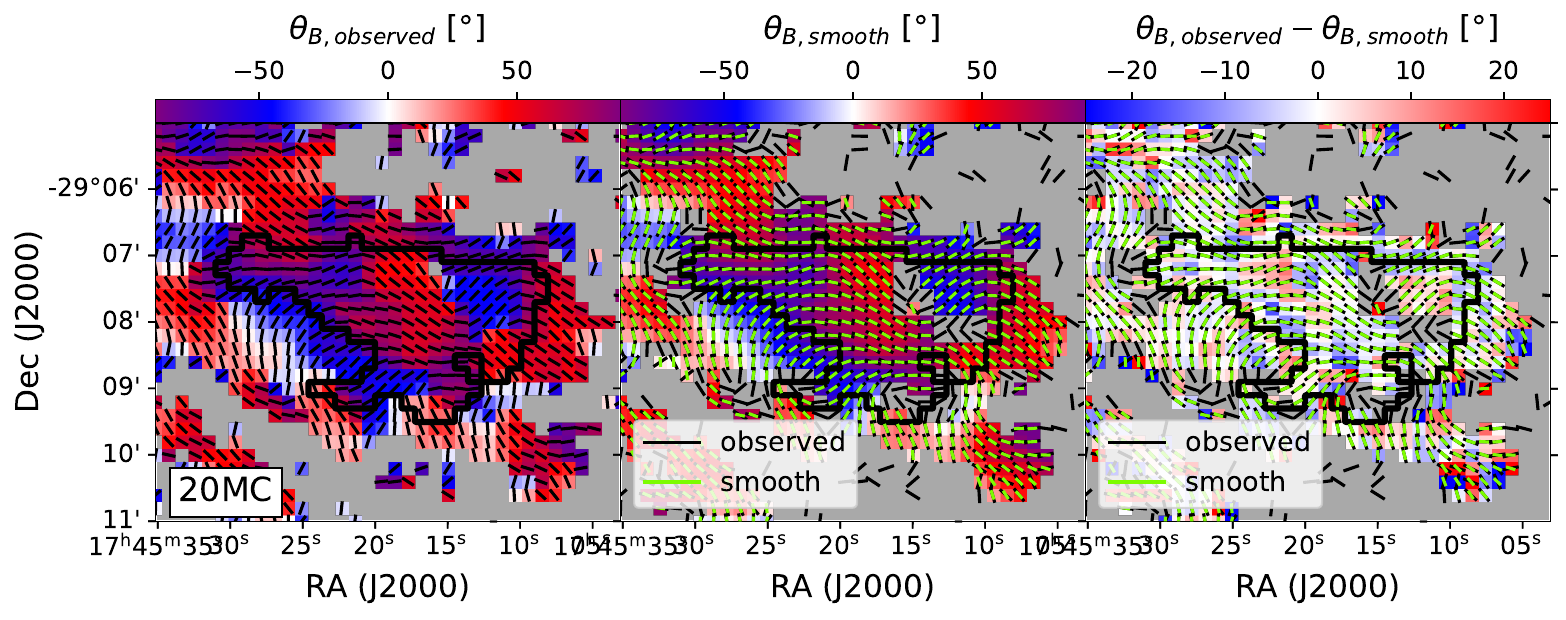}
	\end{subfigure}
	\begin{subfigure}{}
	   \includegraphics[width=0.9\linewidth]{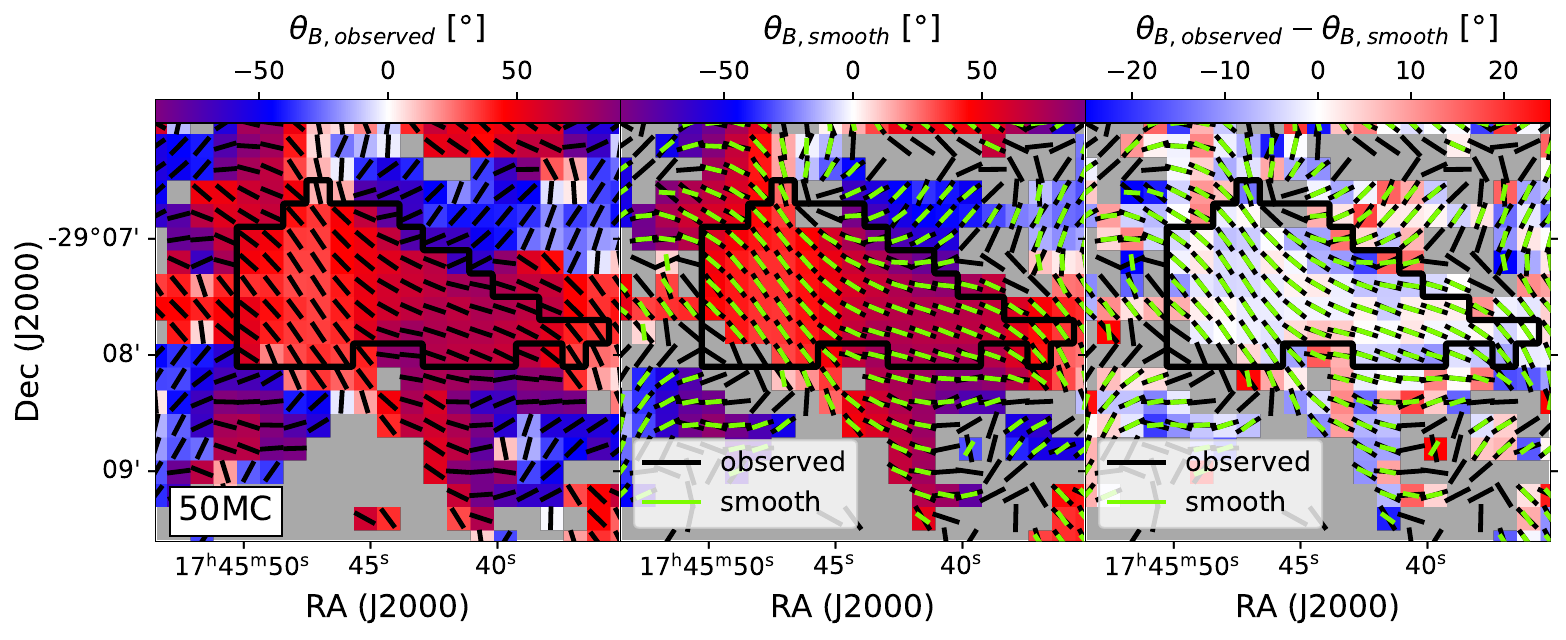}
	\end{subfigure}
	\begin{subfigure}{}
	   \includegraphics[width=0.9\linewidth]{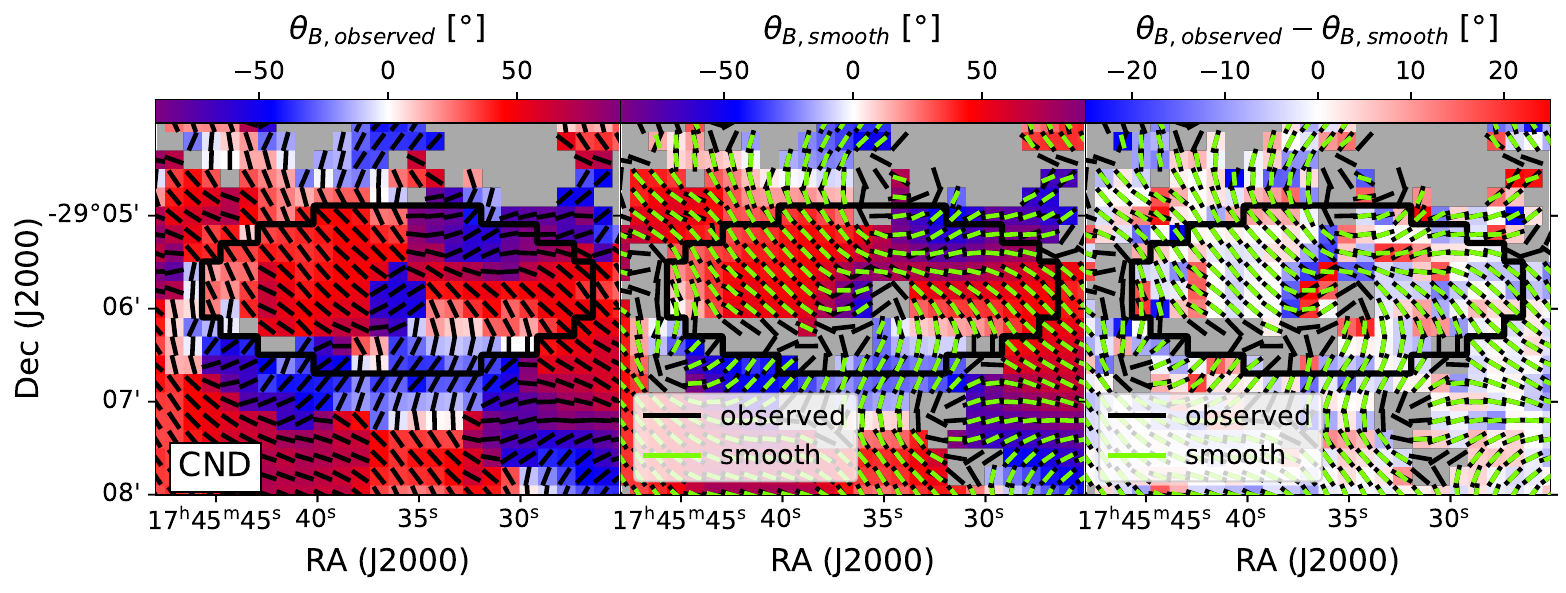}
	\end{subfigure}
	\caption{Same as Figure \ref{fig:smoothPA_450} but for the 850\,$\rm \upmu m$ data.}
	\label{fig:smoothPA_850}
\end{figure*}

We utilize the ``unsharp masking" method, as introduced by \citet{2017ApJ...846..122P} to calculate the angular dispersion of the magnetic fields for the turbulent component at 450\,$\rm \upmu m$ and 850\,$\rm \upmu m$.
They determined the large-scale magnetic field morphology by smoothing the position angles using a boxcar average.
The size of the boxcar filter is chosen so that the smoothing length is smaller than the radius of curvature of the magnetic field morphology.
If there is any pair of the polarization position angles in the boxcar with angle difference higher than 90$\degree$, this pixel at the center of the boxcar will be masked out because the magnetic field is too distorted to determine a mean direction (see \citet{2017ApJ...846..122P} for details).
Then, the mean directions are subtracted from the original ones to obtain the angle differences, which are regarded as the turbulent component of the magnetic field morphology.
The angular dispersion is the standard deviation of the angle differences within a specific region.

We select polarization data with a pixel size of 12$''$ at both 450\,$\rm \upmu$m and 850\,$\rm \upmu$m, using the criteria $\rm SNR_I > 10$ and $\rm SNR_{PI} > 3$.
To determine the size of the boxcar filter, we estimate the radii of curvatures, using the following formula suggested by \citet{2012ApJ...747...79K}:
\begin{equation}
    C\equiv\frac{1}{R}=\frac{2}{d}\cos{(\frac{1}{2}[\pi -\Delta PA])},
    \label{eqn:curvature}
\end{equation}
where C is the curvature; R is the radius of curvature; and d and $\Delta PA$ are the distance and the position angle difference between two adjacent pixels, respectively.
After investigating the curvatures of the magnetic field morphology, we adopt 3 pixels ($\sim$1.5\,pc) as the length of the boxcar filter, which is smaller than most of the radii of local curvatures at both 450\,$\rm \upmu$m and 850 $\upmu$m.
We smooth the magnetic field position angles using a $3\times3$ boxcar filter across the entire region, and produce the residual maps.
If the valid data points are less than 4 within the number of boxcar filter, this position will be masked out.
Fig. \ref{fig:smoothPA_450} and Fig. \ref{fig:smoothPA_850} display the results at 450\,$\rm \upmu$m and 850\,$\upmu$m, respectively. 
The pixels that have been masked out during the process are shown in gray.
The line segments and the color maps display the position angles simultaneously.
The angular dispersion is obtained by calculating the standard deviation of the position angle differences within the corresponding boundary.
Table \ref{table:DCF_parameters} shows the angular dispersion in the 20MC, 50MC, and CND at 450\,$\rm \upmu$m and 850\,$\upmu$m.
The uncertainty in the angular dispersion is estimated by propagating the observational uncertainties.

\subsubsection{Velocity Dispersion} \label{subsubsec:velocity_dispersion}

\begin{figure*} [ht!]
\epsscale{1.15}
\plotone{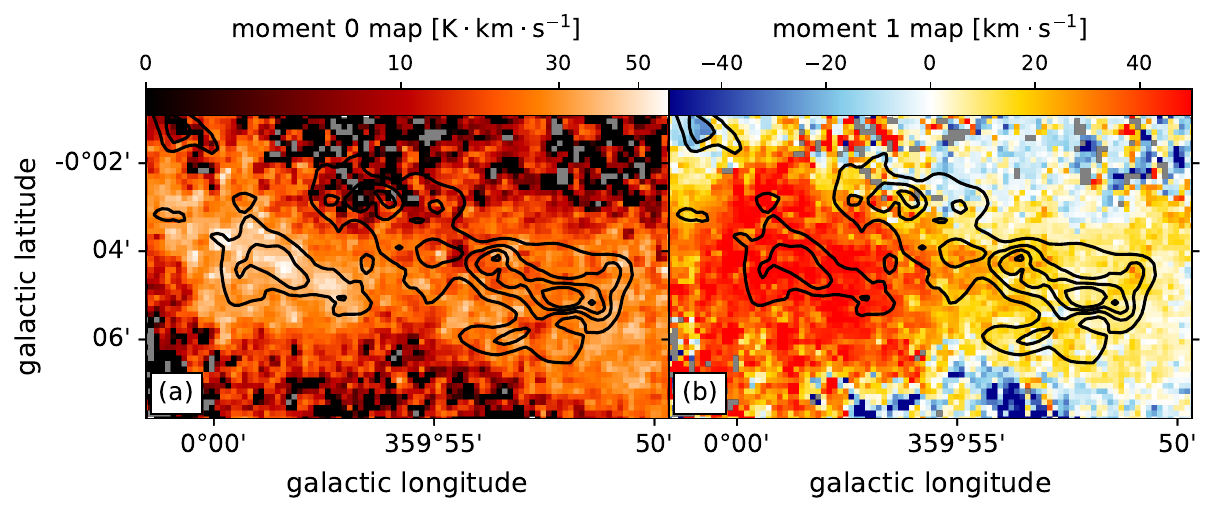}
    \caption{
    The 0th and 1st moment maps of NRO $\rm C^{18}O$ (1-0) with 15$''$ resolution shown in panel (a) and panel (b), respectively. The black contours show the 850\,$\rm \upmu m$ intensities with the levels of [0.8, 1.6, 2.4, 3.2]\,$\rm Jy\,beam^{-1}$.
    \label{fig:moments_nobeyama_c18o}}
\end{figure*}

\begin{figure*} [ht!]
\epsscale{1.05}
\plotone{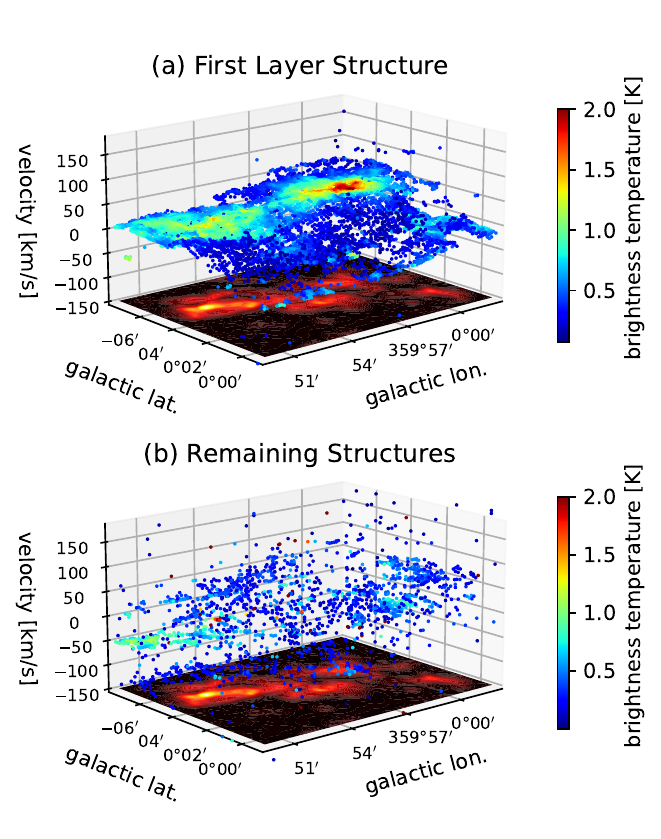}
    \caption{The peak velocities determined from multiple Gaussian fitting using NRO $\rm C^{18}O$ (1-0) data, shown in PPV space.
    Panel (a): The first layer structure.
    Panel (b): The remaining structures, excluding the first layer structure.
    Data points where the fitting amplitude value is less than 5 times the fitting error are excluded.
    The surface on the bottom shows the 850\,$\rm \upmu m$ intensity.
    \label{fig:PPV_3D}}
\end{figure*}

We calculate the average velocity dispersion using the $\rm C^{18}O$ (1-0) observations obtained from NRO with a resolution of 15$''$, assuming that the molecular emission traces approximately the same structures as the submillimeter continuum emission.
In Fig. \ref{fig:moments_nobeyama_c18o}, we mask values below three times the noise level over the spectral line and calculate the 0th and 1st moment maps to investigate the integrated intensity, intensity-weighted velocity, and velocity dispersion, respectively.
In panel (a), we observe bright gas emission within the 20MC and 50MC regions, with fainter emission in the CND.
Panel (b) shows a gradual change in velocity.
The $\rm C^{18}O$ (1-0) spectra, however, exhibit multiple emission peaks in some regions, leading to overestimations of the velocity dispersion derived from the 2nd moment map.
We interpret that the multiple components of the spectrum as being caused by the several structures along the LOS that have different gas dynamics with distinct mean velocity and velocities dispersions.

Accordingly, we fit the spectra with multiple Gaussian profiles to determine the brightness, mean velocity, and velocity dispersion of each emission peak.
The goodness of the fitting is judged by the chi-squared test.
We perform the spectral line fitting pixel by pixel.
Each spectrum is fitted first by a single Gaussian profile and the reduced chi-squared value is calculated.
If the reduced chi-squared value is greater than 2, we add one more Gaussian profile.
This process iterates until the reduced chi-squared value is lower than 2 or the number of Gaussian profiles exceeds 6.
In the event those criteria cannot be met, we assume the spectrum is too noisy to be analyzed.

Subsequently, we analyze the fitting results in the position-position-velocity (PPV) space and attempt to distinguish different gas structures.
We assume that the emission peaks of the spectra correspond to distinct gas structures along the LOS.
Firstly, we sort the fitting results into different groups based on the number of components.
One-component spectra account for $\sim$50\%, $\sim$50\%, and $\sim$70\% of the area within the boundaries of 20MC, 50MC, and CND, respectively.
These components in PPV space exhibit smooth velocity variations, suggesting a similar kinematic structure along the line of sight, which we define as the ``first layer structure''.

Then, we investigate spectra with several components adjacent to pixels with first layer structure by calculating the velocity differences between these components and the component of the first layer structure.
The component with the minimum velocity difference is considered to connect to the first layer structure in the PPV space, if the velocity difference is smaller than the full-width-half-maximum (FWHM) of the component of first layer structure.
We continue iterating this process until no component is classified as part of the first layer structure.
As a result, all components are separated into two groups: the first layer structure and the remaining structures.
The classification results are shown in Fig. \ref{fig:PPV_3D}.
In panel (a) of Fig. \ref{fig:PPV_3D}, the first layer structure covers the regions of the 20MC, 50MC, and CND. 
We can further investigate other gas structures using the remaining components (Panel (b) in Fig. \ref{fig:PPV_3D}), but these are too fragmented in PPV space to define coherent gas structures.

\begin{figure*} [ht!]
\epsscale{1.15}
\plotone{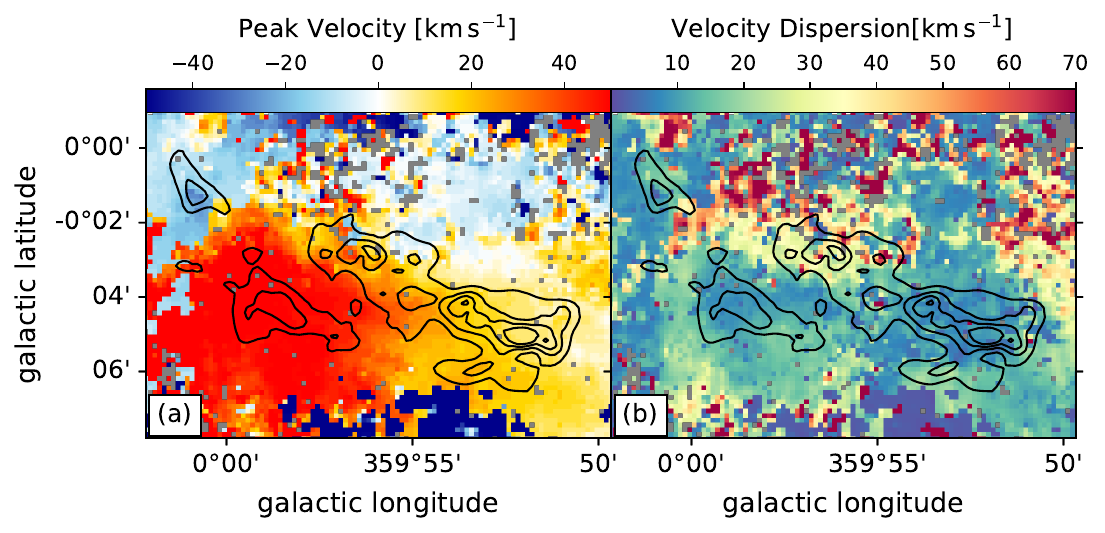}
    \caption{The velocity dispersion determined from $\rm C^{18}O$ (1-0) emission line obtained from NRO.
    Panel (a): the second moment map.
    Panel (b): the velocity dispersion map from the first layer structure in the PPV space.
    Panel (c): the number of the components determined by multiple Gaussian fitting.
    The contours show the 850\,$\rm \upmu m$ intensities with the levels of [0.8, 1.6, 2.4, 3.2]\,$\rm Jy\,beam^{-1}$.
    \label{fig:NRO_vs_fitting}}
\end{figure*}

We adopt the first layer structure in PPV space as indicative of the typical gas motion.
Fig. \ref{fig:NRO_vs_fitting} shows the comparison between the velocity dispersion maps determined by the second moment map (panel (a)) and our analysis (panel (b)).
The velocity dispersions in panel (b) are smaller than those in panel (a) and exhibit more coherence due to the elimination of minor components.
We estimate the non-thermal velocity dispersion by subtracting the thermal component from the $\rm C^{18}O$ velocity dispersion using the relation
\begin{equation}
    \sigma^2_\nu=\sigma^2_{\nu, C^{18} O}-\frac{k_BT}{m_{C^{18}O}},
    \label{eqn:nonth_vd}
\end{equation}
where $\sigma_\nu$ is the non-thermal gas velocity dispersion, $\sigma_{\nu, C^{18} O}$ is the velocity dispersion of the $\rm C^{18}O$ (1-0), and $m_{\rm C^{18}O}$ is the mass of the $\rm C^{18} O$ molecule.
The temperature is taken from the dust temperature shown in Fig. \ref{fig:beta_T} under the assumption the dust and gas are in thermal equilibrium.
By using the boundaries defined in Sect. \ref{subsubsec:volume_density} for the 20MC, 50MC, and CND, we estimate the average velocity dispersion within these regions, and present the results in Table \ref{table:DCF_parameters}.
The uncertainties in the velocity dispersions are estimated by propagating the observational uncertainties and the 1$\sigma$ errors from the spectrum fitting.

The $\sigma_v$ in the CND is significantly larger than that in the 20MC and 50MC. 
\citet{2024MNRAS.531.5012A} obtained a velocity dispersion in the CND that is approximately one-third of our value by decomposing the CHIMPS2 $\rm ^{12}CO$ (3-2) spectra using more Gaussian components.
These $\rm ^{12}CO$ (3-2) observations are brighter than the $\rm C^{18}O$ (1-0) emissions measured by NRO.
We attempted to decompose the CHIMPS2 $\rm ^{12}CO$ (3-2) spectra but encountered degeneracy issues and possible absorption in the spectra. 
Therefore, we continue to use the CHIMPS2 $\rm C^{18}O$ (1-0) as the tracer to determine $\sigma_v$, while cautioning that the decomposition process has inherent limitations.

\subsubsection{Results of DCF Method} \label{subsubsec:DCF_results}

The average values and the corresponding uncertainties of the $n(H_2)$, $\sigma_{\theta}$, and $\sigma_v$ within the structures boundaries of 20MC, 50MC, and CND are shown in Table \ref{table:DCF_parameters} (Note that the boundaries determined using 450\,$\rm \upmu m$ and 850\,$\rm \upmu m$ intensity maps are different).
We use these results to estimate the POS magnetic field strengths and infer the statistical three-dimensional magnetic field strengths using Eq. \eqref{eqn:DCF_B_Pattle} and \eqref{eqn:DCF_B_Crutcher}, respectively.
The POS magnetic field strengths derived from the 450\,$\rm \upmu$m data are 1.0$\rm\pm0.5$\,mG, 1.0$\rm\pm0.5$\,mG, and 2.2$\rm\pm1.2$\,mG, while those from the 850\,$\rm \upmu$m data are 0.7$\rm\pm0.4$\,mG, 0.8$\rm\pm0.4$\,mG, and 1.9$\rm\pm1.1$\,mG in the 20MC, 50MC, and CND regions, respectively.
The magnetic field strength in the CND is approximately twice that estimated in the 20MC and 50MC.
This higher estimate is mainly attributed to the significantly larger $\sigma_v$, which is related to the decomposition process as mentioned in Sect. \ref{subsubsec:velocity_dispersion}. 
It is important to note that our magnetic field strength estimations, especially in the CND, may be overestimated due to potential limitations in the spectral decomposition process.

The difference in magnetic field strengths derived from the 450\,$\rm \upmu m$ and 850\,$\rm \upmu m$ data is due to the discrepancy in $n(H_2)$, which is largely influenced by variations in the observed intensity.
As discussed in Sect. \ref{subsubsec:volume_density}, the intensities at 450\,$\rm \upmu m$ and 850\,$\rm \upmu m$ are not consistent with a common SED, suggesting that they may not trace the same dust structures along the LOS. 
This discrepancy is likely due to missing flux issues (see Appendix \ref{appendix:CO_removal} for details). 
Consequently, we interpret the magnetic field strengths derived from different wavelengths as probing slightly different regions along the LOS.

\citet{2024ApJ...962...39L} concluded that self-gravity and turbulence play a more dominant role in star formation, based on the weak correlation between  magnetic field strength and SFR.
However, we believe this weak correlation does not imply star formation is unaffected by magnetic fields; rather, it suggests the SFR is not linearly correlated with magnetic field strength.
Star-forming processes should not depend solely on the influence of the magnetic field.
Therefore, comparing the energy budgets of the magnetic field, gravity, turbulence, and thermal energy is essential.



\begin{table*}[ht!]
\centering
\caption{Values of number density, FWHM velocity dispersion, angular dispersion, and dust temperature used to estimate the magnetic field strengths using the DCF method within the regions of 20MC, 50MC, and CND.
}
\begin{tabular}{lcccccccc} 
\hline
\hline 
\\[-1.4em]
    {} &  
    \multicolumn{2}{b{2.2cm}}{\centering{$n(H_2)$ [$cm^{-3}$]}} & 
    \multicolumn{2}{b{2.2cm}}{\centering{$\sigma_v$ [
$km\cdot s^{-1}$]}}&
    \multicolumn{2}{b{2.2cm}}{\centering{$\sigma_\theta$ [$\degree$]}} & 
    \multicolumn{2}{b{2.2cm}}{\centering{T [$K$]}}
    \\[0.1em]    
\cmidrule{2-9}
    \multicolumn{1}{b{1.1cm}}{
    \centering{Region}} & 
    {\centering$\rm 450\,\upmu m$} &  {\centering{$\rm 850\,\upmu m$}} & 
    {\centering$\rm 450\,\upmu m$} &  {\centering{$\rm 850\,\upmu m$}} &  
    {\centering$\rm 450\,\upmu m$} &  {\centering{$\rm 850\,\upmu m$}} & 
    {\centering$\rm 450\,\upmu m$} &  {\centering{$\rm 850\,\upmu m$}} 
    \\[0.5em]
\hline
\\[-0.5em]
\multicolumn{1}{b{1.1cm}}{\centering{20MC}} & 4437 & 2141 & 11.3 & 10.3 & 9.3  & 8.4  & 18.6 & 18.2\\
\multicolumn{1}{b{1.1cm}}{\centering{}}     & (572) & (235) & (0.7)& (0.6)& (0.4)& (0.4)& (1.2)& (1.2) \\[1.5ex]
\multicolumn{1}{b{1.1cm}}{\centering 50MC}  & 2624 & 1335 & 10.3 & 10.2 & 6.7  & 5.7  & 23.0 & 22.9\\
\multicolumn{1}{b{1.1cm}}{\centering {}}    & (471) & (192)  & (0.4)& (0.5)& (0.8)& (0.7)&(1.9)& (1.9) \\[1.5ex]
\multicolumn{1}{b{1.1cm}}{\centering CND}   & 3464 & 2897 & 31.8 & 31.7 & 10.4  & 10.9  & 52.7 & 52.7\\
                                            & (724) & (616) & (6.6)& (6.7)& (0.7)& (0.4)& (36.1)& (36.1) 
\\[2.2ex]
\hline
\hline
\multicolumn{9}{b{11cm}}{Note. Values in parentheses indicate the corresponding uncertainties.} \\
\multicolumn{9}{b{11cm}}{Note. Uncertainties are calculated by propagating the uncertainties of the observational data, $\rm1\sigma$ errors from the HiGAL SED fitting, and $\rm1\sigma$ errors from the $\rm C^{18}O$ (1-0) spectrum fitting.
}
\\
\end{tabular}
\label{table:DCF_parameters}
\end{table*}

\begin{table*}[]
    \caption{Magnetic field strength and Energy Budget}
    \setlength{\tabcolsep}{1.mm}{
    \begin{tabular}{lcccccccccc}\hline\hline \\[-1em]
        {} &  
        \multicolumn{2}{b{2.2cm}}{\centering $\rm B_{POS}\,[mG]$} &  
        \multicolumn{2}{b{2.2cm}}{\centering $\rm B_{tot}\,[mG]$} &
        \multicolumn{2}{b{2.2cm}}{\centering $\rm M_{A}$} &
        \multicolumn{2}{b{2.2cm}}{\centering $\rm \overline{(M/\Phi)}/(M/\Phi)_{crit}$} &
        \multicolumn{2}{b{2.2cm}}{\centering $\rm \beta_{plasma}\,[\times10^{-4}]$}\\[0.5em]
    \cmidrule{2-11}
    \multicolumn{1}{b{1.1cm}}{\centering Region} & $\rm 450\,\upmu m$ &  $\rm 850\,\upmu m$ & $\rm 450\,\upmu m$ & $\rm 850\,\upmu m$ & $\rm 450\,\upmu m$ &  $\rm 850\,\upmu m$ & $\rm 450\,\upmu m$ & $\rm 850\,\upmu m$ & $\rm 450\,\upmu m$ & $\rm 850\,\upmu m$\\[0.5em]
    \hline\\[-0.5em]
    \multicolumn{1}{b{1.1cm}}{\centering 20MC} & 1.0 & 0.7 & 1.3 & 0.9 & 0.44 & 0.40 & 0.18 & 0.13 & 1.8 & 1.7\\
    \multicolumn{1}{b{1.1cm}}{\centering } & (0.5)  & (0.4) & (0.6) & (0.4) & (0.02)  & (0.02) & (0.09) & (0.07) & (1.8) & (1.8)\\[1.5ex]
    \multicolumn{1}{b{1.1cm}}{\centering 50MC} & 1.0 & 0.8 & 1.2 & 1.0 & 0.32 & 0.27 & 0.08 & 0.05 & 1.4 & 1.0\\
    \multicolumn{1}{b{1.1cm}}{\centering } & (0.5) & (0.4) & (0.6) & (0.5) & (0.04)  & (0.03) & (0.05) & (0.03) & (1.4) & (1.1)\\[1.5ex]
    \multicolumn{1}{b{1.1cm}}{\centering CND} & 2.2 & 1.9 & 2.8 & 2.5 & 0.50 & 0.52 & $<$0.01 & $<$0.01 & 0.8 & 0.9\\
    \multicolumn{1}{b{1.1cm}}{\centering } & (1.2) & (1.1) & (1.6) & (1.4) & (0.04)  & (0.02) & ($<$0.01) & ($<$0.01) & (0.1) & (0.1)\\[2.2ex]
    \hline\hline
    \multicolumn{11}{b{9.9cm}}{Note. Values in parentheses indicate the corresponding uncertainties.} \\
    \end{tabular}}
    \label{table:B_strength}
\end{table*}

\subsection{Alfvén Mach Number} \label{subsec:Mach_number}

The relative importance between the magnetic field and the non-thermal motions can be described by the Alfvén Mach number ($M_A$), which is a crucial factor in the cloud evolution models.
The definition of the Alfvén Mach number is 
\begin{equation}
    M_A=\sqrt{3}\frac{\sigma_{NT}}{V_A},
    \label{eqn:M_A}
\end{equation}
where the Alfvén speed is
\begin{equation}
    V_A=\frac{B}{\sqrt{4\pi \rho}}.
    \label{eqn:V_A}
\end{equation}
Here, $\sigma_{NT}$ represents the non-thermal velocity dispersion which we assume is due to turbulent perturbation, and $V_A$ is the Alfvén velocity.
By incorporating Eq. \eqref{eqn:DCF_B_Pattle}, \eqref{eqn:DCF_B_Crutcher} and \eqref{eqn:V_A} into Eq. \eqref{eqn:M_A}, we obtain:
\begin{equation}
    M_A\approx\frac{\sqrt{3}\pi}{4}\frac{\sigma_{\theta}}{Q'}.
    \label{eqn:M_A_from_sigma_theta}
\end{equation}
This formula implies that the Alfvén Mach number can be estimated directly from the angular dispersion of magnetic field orientations.

We estimate the Alfvén Mach number for the 20MC, 50MC, and CND using Eq. \eqref{eqn:M_A_from_sigma_theta}, and present the results in Table \ref{table:B_strength}.
The values are within the range of 0.3 to 0.5, indicating the non-thermal motions are sub-Alfvénic.

\subsection{Thermal to Magnetic Energy} \label{subsec:ThermalEnergy}
To determine the importance of the magnetic field relative to the thermal energy, we adopt the standard $\beta_{plasma}$, the gas-to-magnetic pressure ratio, defined by
\begin{equation}
    \beta_{plasma}=\frac{P_{gas}}{P_{mag}}=\frac{2c_s^2}{V_A^2},
    \label{eqn:plasma_beta}
\end{equation}
where 
\begin{equation}
    c_s=\sqrt{\frac{k_BT}{\mu_{H_2}m_H}},
\end{equation}
Here, $\rm P_{gas}$ is the gas pressure, $\rm P_{mag}$ is the magnetic pressure, and $\rm c_s$ is the sound speed.
We consider the dust temperature to calculate the sound speed under the assumption of the gas and dust are in thermal equilibrium.
If $\beta_{plasma}$ is smaller than 1, it implies a magnetically dominated regime.

We calculate $\beta_{plasma}$ for the 20MC, 50MC, and CND, and the results are presented in Table \ref{table:B_strength}.
All $\beta_{plasma}$ values are much smaller than 1, on the order of $\sim 10^{-4}$, suggesting magnetic field dominance.

\begin{figure} [ht!]
\epsscale{1.15}
\plotone{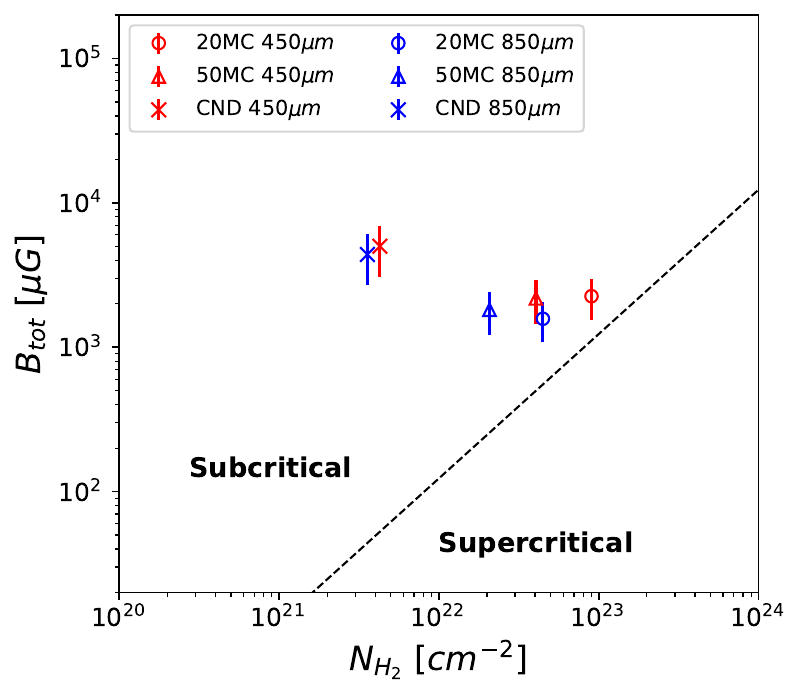}
    \caption{The relation between the total magnetic field strength and the column density of the hydrogen molecule at 450\,$\rm \upmu m$ and 850\,$\rm \upmu m$.
    The line is for the criterion of $M/\Phi = (M/\Phi)_{crit}$.
    \label{fig:B_N_relation}}
\end{figure}

\subsection{Mass-to-Flux Ratio} \label{subsec:MFratio}

The significance of magnetic fields in the relation to gravity is crucial when considering gravitational instability, which can be assessed through the mass-to-flux ratio.
\citet{1978PASJ...30..671N} proposed a critical mass-to-flux ratio, $(M/\Phi)_{crit}=(2\pi \sqrt{G})^{-1}$, for an isothermal gaseous disk threaded perpendicularly by a uniform, frozen-in magnetic field using linear perturbation analysis.
It is common to express the mass-to-flux ratio in the unit of the critical value, which is defined as $\lambda\equiv(M/\Phi_B)_{obs}/(M/\Phi_B)_{crit}$.
The molecular cloud is gravitationally stable when $\lambda<1$, regarded as ``magnetically subcritical".
In contrast, if $\lambda>1$, named as ``magnetically supercritical", the cloud will be unstable and start to collapse toward the center.
The mass-to-flux ratio can be determined from observations using the column density $N$ and the magnetic field strength $|B|$:
\begin{equation}
    \begin{aligned}
        \lambda&=\frac{(M/\Phi)_{obs}}{(M/\Phi)_{crit}}=\frac{(mNA)/(|B|A)}{(2\pi \sqrt{G})^{-1}}\\ &\approx 7.6\times 10^{-21}\frac{N(H_2)}{|B|}.
    \end{aligned}
\label{mass-to-flux-ratio}
\end{equation}

\begin{figure*} [ht!]
\epsscale{1.1}
\plotone{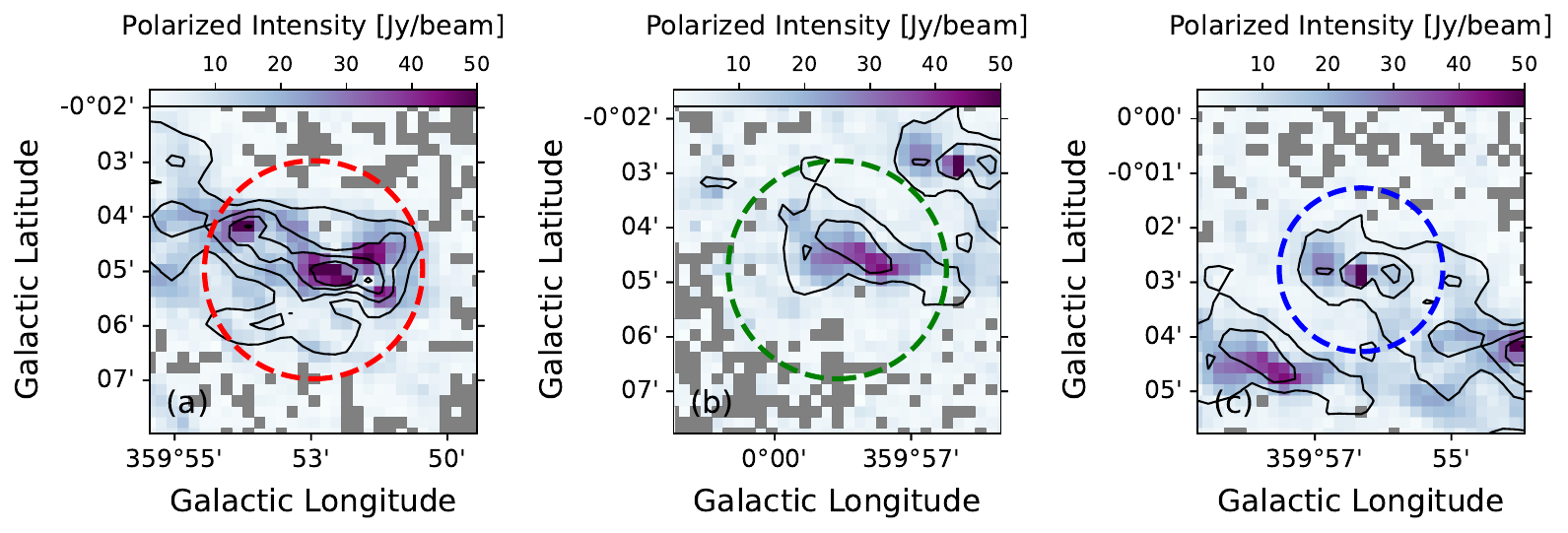}
    \caption{
    Regions with diameters of 4$'$, 4$'$, and 3$'$ are shown by dashed circles around (a) 20MC, (b) 50MC, and (c) CND, overlaid with the polarization intensity at 850\,$\rm \upmu m$.
    The contours show the CO-removed
    850\,$\rm \upmu m$ intensities with the values of [0.5, 1, 2, 3, 4]\,$\rm Jy\, beam^{-1}$.
    \label{fig:diff_regions_for_RiceanMeanModelFitting}}
\end{figure*}

Fig. \ref{fig:B_N_relation} displays the relationship between the total magnetic field strength and hydrogen molecule column density.
The dashed line represents the critical value for the mass-to-flux ratio. 
Dots located above the dashed line indicate a subcritical condition, while dots below the line signify a supercritical condition. 
Based on the estimations, the $\lambda$ values for the 20MC, 50MC, and CND are subcritical.

\citet{2004mim..proc..123C} showed that a statistical average mass-to-flux ratio for the strongly magnetized case is
\begin{equation}
    \begin{aligned}
        \overline{\left(\frac{M}{\Phi}\right)}&=\int_{0}^{\pi/2}\frac{M\ cos\theta}{\Phi/sin\theta}sin\theta\    d\theta\\ &=\frac{1}{3}(\frac{M}{\Phi})_{obs},
    \end{aligned}
    \label{eqn:MBratio_approx}
\end{equation}
where clouds have a disk-like morphology.
We calculate the mass-to-flux ratio using Eq. \eqref{eqn:MBratio_approx} normalized to the critical value using this statistical average and show the results in Table \ref{table:B_strength}.
The average mass-to-flux ratios are consistently below 1 within the regions of 20MC, 50MC, and CND, suggesting that these structures are stable and effectively supported by the magnetic field.

\subsection{Grain Alignment Efficiency} \label{subsec:align}

\begin{figure*}
    \centering
    \begin{subfigure}{}
        \includegraphics[width=0.8\textwidth]{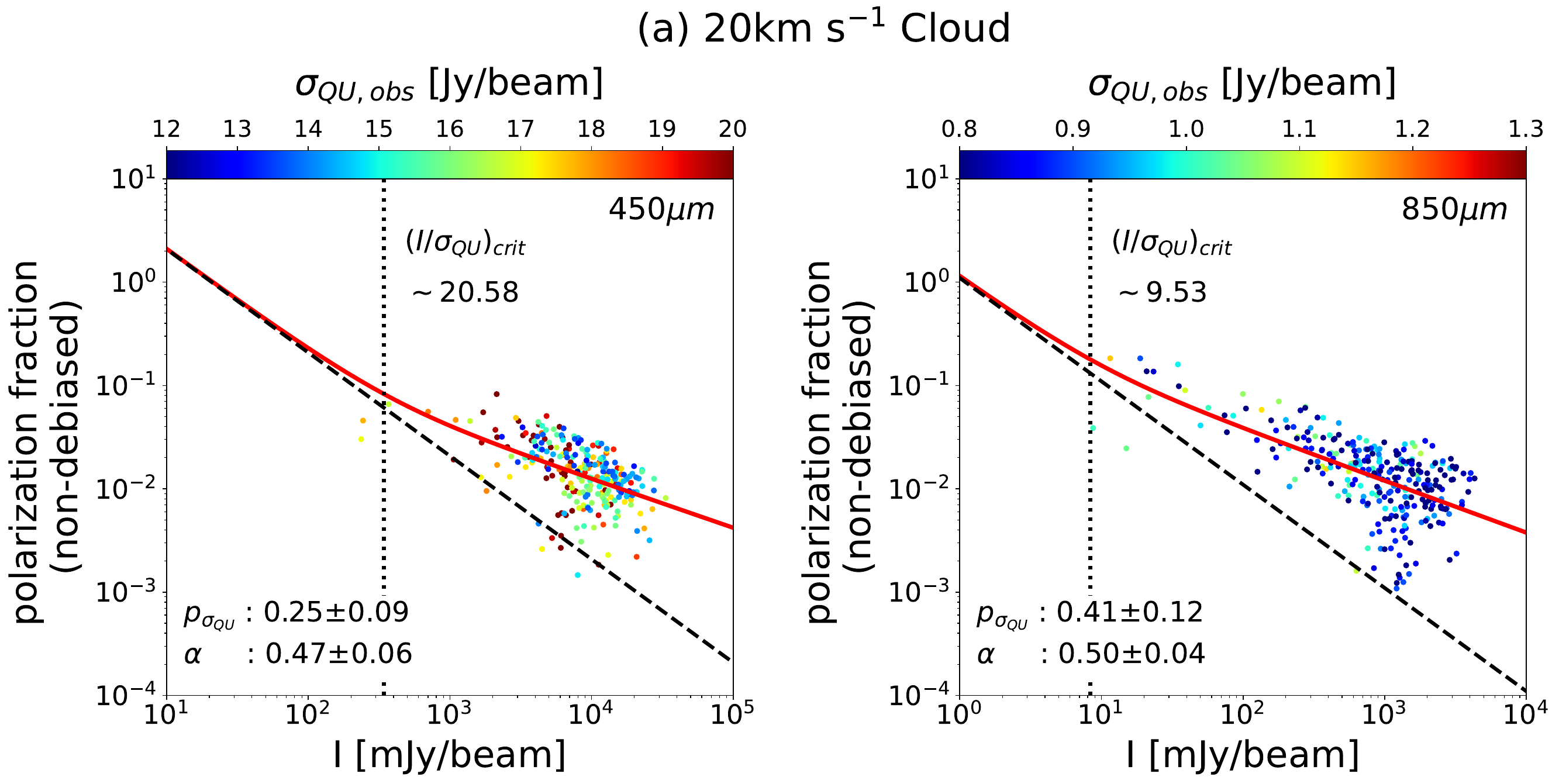}
        \label{fig:first}
    \end{subfigure}
    \begin{subfigure}{}
        \includegraphics[width=0.8\textwidth]{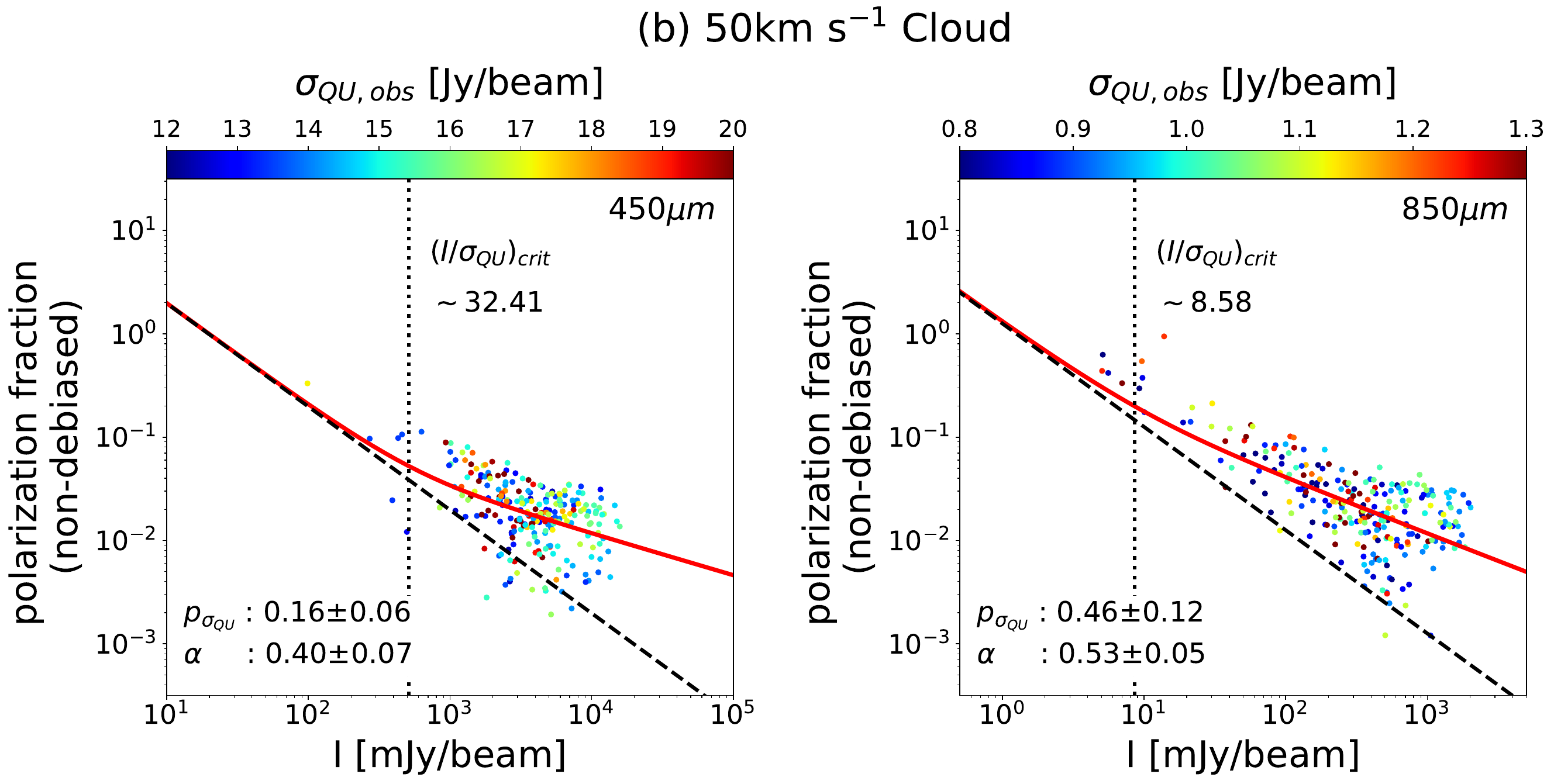}
        \label{fig:second}
    \end{subfigure}
    \begin{subfigure}{}
        \includegraphics[width=0.8\textwidth]{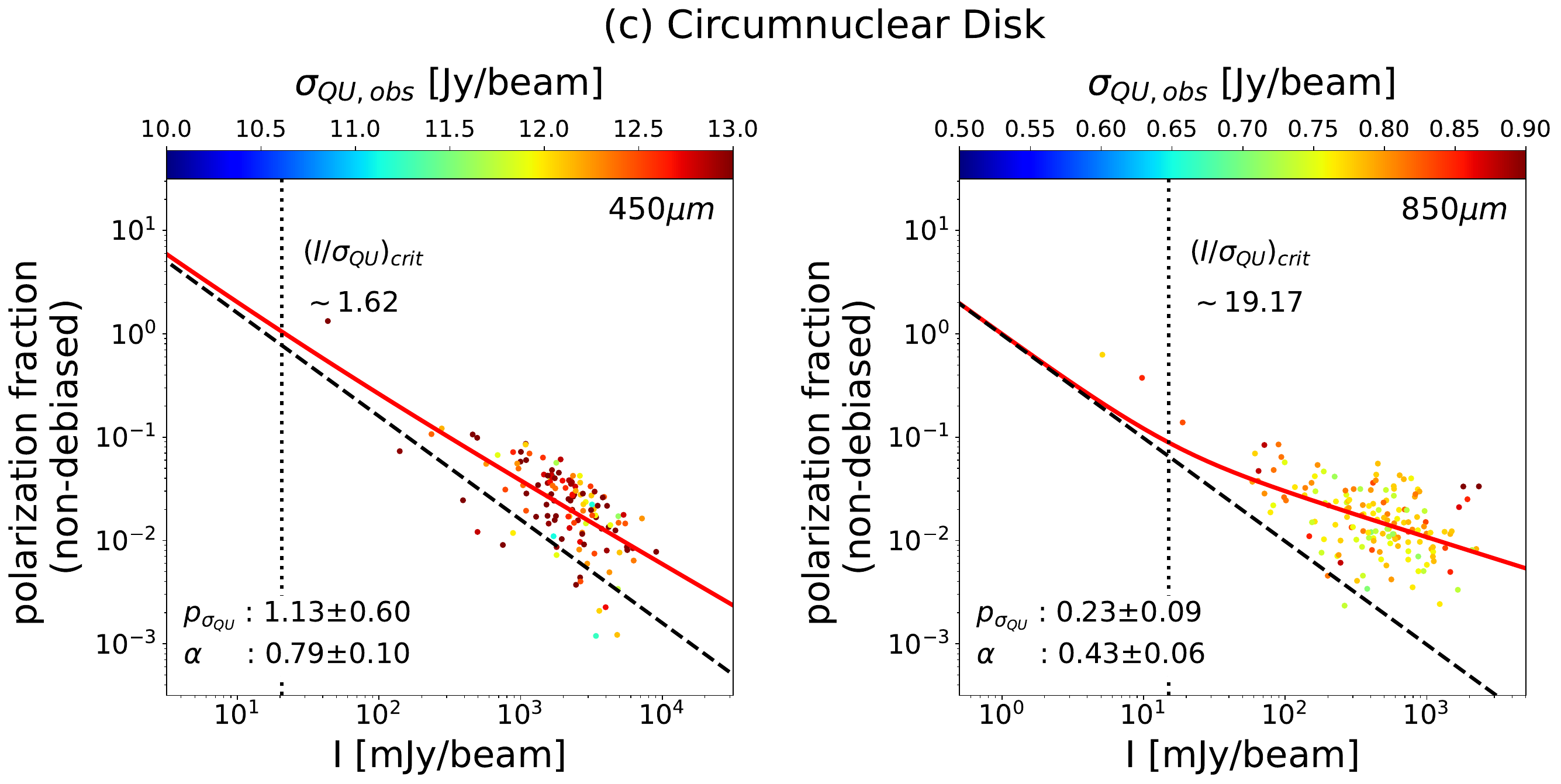}
        \label{fig:third}
    \end{subfigure}
    \caption{The biased polarization fraction against the Stokes I for (a) 20MC, (b) 50MC, and (c) CND 
    The data are extracted within the boundaries shown in Fig. \ref{fig:diff_regions_for_RiceanMeanModelFitting}.
    The red solid lines show the fitting results using Ricean-mean model.
    The black dashed lines present the low-SNR limit approximation of Eq. \eqref{eqn:Ricean_mean_model}.
    The black dotted lines indicate the critical value derived from Eq. \eqref{eqn:cri_IDQU}.
    }
    \label{fig:PIvsI_and_Ricean_model_diff_regions}
\end{figure*}

The power law index $\alpha$ of the dependence of the polarization fraction on the total intensity quantifies the grain alignment efficiency, $p\propto I^{-\alpha}$ \citep[][]{2015ARA&A..53..501A}, assuming that LOS depolarization is negligible.
$\alpha$ ranges from 0 to 1.
In molecular clouds, the power-law index potentially indicates the alignment efficiency between the dust grains and the magnetic field.
If $\alpha \sim 1$, it implies that dust grains are not aligned with any specific direction or are influenced by the disorganized component of the magnetic fields. 
By contrast, if $\alpha \sim 0$, it indicates that the dust grains are well aligned with each other.
Because $\alpha$ may be overestimated due to the non-Gaussian noise characteristics of the polarization fraction, \citet{2019ApJ...880...27P} proposed a ``Ricean-mean model", which can recover $\alpha$ using the mean of the Rice distribution by fitting the p-I relation with the biased polarization fraction $p'$, as follows: 
\begin{small}
\begin{equation}
    p' = \sqrt{\frac{\pi}{2}}
    \left( \frac{I}{\sigma_{QU}} \right)^{-1}
    \mathcal{L}_{\frac{1}{2}}\left[ -\frac{p_{\sigma_{QU}}^2}{2}\left(\frac{I}{\sigma_{QU}}\right)^{2(1-\alpha)} \right].
    \label{eqn:Ricean_mean_model}
\end{equation}
\end{small}
Here, $\sigma_{QU}$ is the average noise in Stokes Q and U from the associated variance $V_Q$ and $V_U$, $p_{\sigma_{QU}}$ is the polarization fraction at the noise level, and $\mathcal{L}_{\frac{1}{2}}$ is a Laguerre polynomial of order $1/2$. 
They assume that the observed data can be characterized by a single RMS noise value, $\left<\sigma_{QU}\right>$,
\begin{equation}
    \left<\sigma_{QU}\right>=\frac{1}{2N}\sum\limits_{i=1}^N \sigma_{QU,i}=\frac{1}{2N}\sum\limits_{i=1}^N\left(\sqrt{V_{Q,i}}+\sqrt{V_{U,i}}\right),
    \label{eqn:sigma_QU}
\end{equation}
The underlying power-law index, which indicates the alignment efficiency, is revealed when
\begin{equation}
    \frac{I}{\sigma_{QU}}>\left(\frac{I}{\sigma_{QU}}\right)_{crit}=\left( \frac{1}{p_{\sigma_{QU}}}\sqrt{\frac{\pi}{2}}\right)^{\frac{1}{1-\alpha}}.
    \label{eqn:cri_IDQU}
\end{equation}
When $I/\sigma_{QU}$ is lower than the critical value, the power-law index tends to approach -1. 
For a detailed verification, see \citet{2019ApJ...880...27P}.

To investigate the grain alignment efficiencies in the 20MC, 50MC, and CND, we extract the Stokes I, Q, and U values within circular areas of 4$'$ diameter centered on the 20MC and 50MC, and within a 3$'$ diameter region centered on the CND, as shown in Fig. \ref{fig:diff_regions_for_RiceanMeanModelFitting}.
Using these values, we calculate the biased polarization fractions ($p'=\sqrt{Q^2+U^2}/I$) and estimate the average noise levels $\left<\sigma_{QU}\right>$ in each area.
We then fit the data points of Stokes I against the biased polarization fraction using the Ricean-mean model.
Fig. \ref{fig:PIvsI_and_Ricean_model_diff_regions} presents the scatter plot of Stokes I versus the biased polarization fraction, with $\sigma_{QU}$ indicated by different colors.
The red solid lines indicate the best-fit Ricean-mean model, the black dashed lines represent the low-SNR limit approximation, $p'=\sqrt{\pi/2}\left< I/\sigma_{QU} \right>^{-1}$, and the black dotted lines denote the critical value as Eq. \eqref{eqn:cri_IDQU}.
The data points in Fig. \ref{fig:PIvsI_and_Ricean_model_diff_regions} are well characterized by $\left<\sigma_{QU}\right>$ and well fitted by Ricean-mean model.

The fitting results are summarized in Table \ref{table:RMMfitting}.
Except for the polarization observations at 450\,$\rm \upmu m$ towards the CND, the power law index $\alpha$ ranges from 0.4 to 0.6, indicating a moderately ordered alignment of the dust grains.
In the case of 450\,$\rm \upmu$m polarization in the CND region, we obtain a higher power-law index, exceeding 0.79, which implies poor alignment between the magnetic fields and the dust grains.

\begin{table}[ht!]
\centering
\caption{Fitting results using Ricean-mean model for 20MC, 50MC, and CND.}
\begin{tabular}{rccc} 
\hline
\hline 
\\[-1.4em]
    \multicolumn{1}{b{1cm}}{} & 
    \multicolumn{1}{b{2cm}}{} &
    \multicolumn{2}{b{4cm}}{\centering{Ricean-mean Model}}\\
\cmidrule{3-4}
\\[-1.8em]
    \multicolumn{1}{b{1cm}}{\centering{Region}} &  
    \multicolumn{1}{b{2cm}}{\centering{$\left<\sigma_{QU}\right>$}} & 
    \multicolumn{1}{b{2cm}}{\centering{$\alpha$}}&
    \multicolumn{1}{b{2cm}}{\centering{$p_{\sigma_{QU}}$}}\\[0.3em]
\hline 
    \multicolumn{4}{b{7cm}}{\centering{\ \ \ \ \ \ \  450\,$\rm \upmu m$}}\\[0.3em]
\hline
    \multicolumn{1}{b{1cm}}{\centering{20MC}} &  
    \multicolumn{1}{b{2cm}}{\centering{$16.61$}} & 
    \multicolumn{1}{b{2cm}}{\centering{$0.47\pm0.06$}}&
    \multicolumn{1}{b{2cm}}{\centering{$0.25\pm0.09$}}
    \\[0.3em]
    \multicolumn{1}{b{1cm}}{\centering{50MC}} &  
    \multicolumn{1}{b{2cm}}{\centering{$15.75$}} & 
    \multicolumn{1}{b{2cm}}{\centering{$0.40\pm0.07$}}&
    \multicolumn{1}{b{2cm}}{\centering{$0.16\pm0.06$}}
    \\[0.3em]
    \multicolumn{1}{b{1cm}}{\centering{CND}} &  
    \multicolumn{1}{b{2cm}}{\centering{$12.69$}} & 
    \multicolumn{1}{b{2cm}}{\centering{$0.79\pm0.10$}}&
    \multicolumn{1}{b{2cm}}{\centering{$1.13\pm0.60$}}\\[0.3em]
\hline
    \multicolumn{4}{b{7cm}}{\centering{\ \ \ \ \ \ \ 850\,$\rm \upmu m$}}\\[0.3em]
\hline
    \multicolumn{1}{b{1cm}}{\centering{20MC}} &  
    \multicolumn{1}{b{2cm}}{\centering{$0.88$}} & 
    \multicolumn{1}{b{2cm}}{\centering{$0.50\pm0.04$}}&
    \multicolumn{1}{b{2cm}}{\centering{$0.41\pm0.12$}}
    \\[0.3em]
    \multicolumn{1}{b{1cm}}{\centering{50MC}} &  
    \multicolumn{1}{b{2cm}}{\centering{$1.00$}} & 
    \multicolumn{1}{b{2cm}}{\centering{$0.53\pm0.05$}}&
    \multicolumn{1}{b{2cm}}{\centering{$0.46\pm0.12$}}
    \\[0.3em]
    \multicolumn{1}{b{1cm}}{\centering{CND}} &  
    \multicolumn{1}{b{2cm}}{\centering{$0.78$}} & 
    \multicolumn{1}{b{2cm}}{\centering{$0.43\pm0.06$}}&
    \multicolumn{1}{b{2cm}}{\centering{$0.23\pm0.09$}}\\[0.3em]
\hline
\hline
\\[-1em]
\end{tabular}
\label{table:RMMfitting}
\end{table}

\section{Discussion} \label{sec:discussion}

\subsection{Energy Budget} \label{subsec:energy}
In Sect. \ref{subsec:Mach_number}, \ref{subsec:ThermalEnergy}, and \ref{subsec:MFratio}, we compared the energy contributions between the magnetic field and the turbulence, kinetic energy, and gravity based on the Alfvén Mach Number, $\beta_{plasma}$, and mass-to-flux ratio, respectively, within the regions of the 20MC, 50MC, and CND.
Our results show these regions are sub-Alfvénic.
Furthermore, the $\beta_{plasma}$ values, all below 1 for the observed structures, indicate the dominance of the magnetic pressure over thermal pressure.
Additionally, the mass-to-flux ratios for the 20MC, 50MC, and CND suggest that the magnetic field can effectively counteract gravitational collapse.

The magnetic field strengths of some molecular clouds in the CMZ were measured by previous studies.
\citet{1996ApJ...462L..79C} obtained the LOS magnetic field strength towards the active star-forming cloud, Sgr\,B2, as $\sim$\,0.5\,mG with the VLA using a Zeeman measurement of the H\,{\footnotesize I} line.
They calculated the $\beta_{plasma}$, Alfvén Mach Number, and the mass-to-flux ratio as $8\times 10^{-4}$, 0.4, and 2.6 , respectively.
Their $\beta_{plasma}$ value is similar to our results, with a value much lower than 1.
In Sgr\,B2, however, their mass-to-flux ratio is higher than 1, indicating that the magnetic field there may not be able to counteract gravitational collapse.
This difference could explain why Sgr\,B2 has a higher SFR compared to the 20MC and 50MC.

In addition, \citet{2015ApJ...799...74P} estimated the total magnetic field strength in G0.253+0.016 as approximately $5$\,mG using the DCF method and inferred a $\lambda$ value of roughly $0.6$.
\citet{2016ApJ...832..143F} adjusted the magnetic field strength to 2.2\,mG using an improved average volume density.
There, the magnetic field emerges as a dominant factor in this massive molecular cloud, which is similar in environment to the 20MC and 50MC.
This moderate values of lambda shared by all three clouds may be consistent with the comparatively lower SFRs in G0.253+0.016, the 20MC, and  the 50MC.

The material within the CND is influenced by the gravitational potential of a supermassive black hole and follows a quasi-Keplerian rotation, indicating that gravity is dominant. 
Our results show a subcritical condition for the mass-to-flux ratio in this region, suggesting the magnetic field is more significant.
\citet{2024MNRAS.531.5012A} produced a pixel-by-pixel mass-to-flux ratio map and found a supercritical condition only within the arm-like structures, which are dense regions with low magnetic field strength. 
Our subcritical result may be due to us estimating an average value within the CND region, which contains both mass-to-flux ratios higher and lower than 1.
Additionally, it is also possible the magnetic field strength has been overestimated given the potential limitations in spectral decomposition (see Sect. \ref{subsubsec:DCF_results}).

Consequently, the 20MC, 50MC and the CND exhibit strong turbulence, as evidenced by their high velocity dispersions, a feature supported by previous studies. 
Our findings here reveal a significant magnetic field in these regions, as well, suggesting that the magnetic field dominates the overall energy budget.

\subsection{Magnetized Field and Density Relation} \label{subsec:magnetized_condition}

\begin{figure}
\epsscale{1.15}
\plotone{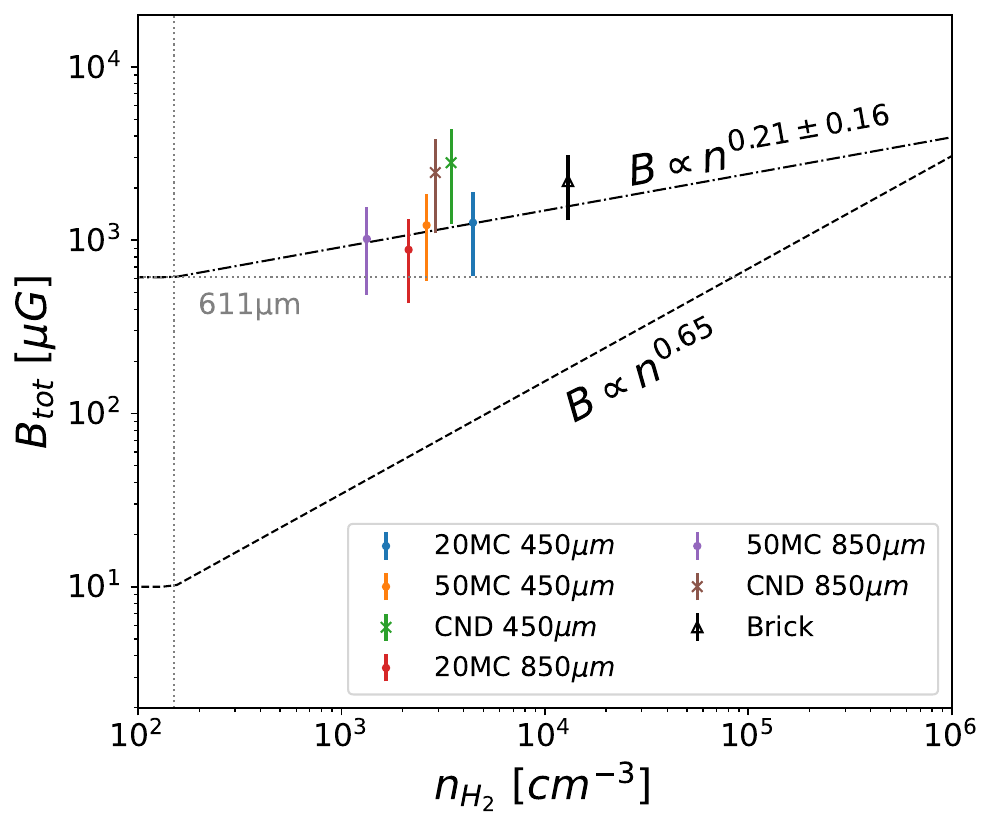}
    \caption{The relation between the total magnetic field strength and the number density of the hydrogen molecule at 450\,$\rm \upmu m$ and 850\,$\rm \upmu m$.
    The dashed line shows the most probable maximum values for the magnetic field strength by the Bayesian analysis \citep[][]{2010ApJ...725..466C}.
    We fit power law index using the data of 20MC, 50MC and Brick \citep[][]{2016ApJ...832..143F} and depict the result by dash-dotted line.
    \label{fig:B_n_relation}}
\end{figure}


The magnetic field strengths in self-gravitating clouds can be enhanced through the contraction of collapsing material when the magnetic fields are initially too weak to resist the collapse of the cloud.
\citet{2010ApJ...725..466C} analyzed the LOS magnetic field strengths $B_{LOS}$ from Zeeman measurements using a Bayesian approach, and suggested a generalized model to describe the relationship between the cloud density and the maximum magnetic field strength in a cloud, as follows:
\begin{equation}
    B_{max}(n)=\left\{
    \begin{aligned}
        &B_0       &,n<n_0\\
        &B_0(\frac{n}{n_0})^{\kappa}&,n>n_0
    \end{aligned}
    \right.
\label{eqn:Crutcher_model}
\end{equation}
where n is the number density of the hydrogen atom ($n_H=2n_{H_2}$).
The critical volume density $n_0$ distinguishes the diffuse interstellar medium and dense molecular clouds.
Their results showed $B_0\approx10\, \upmu G$, $n_{0,H}\approx300\, cm^{-3}$, and $\kappa\approx0.65$.

Assuming that the magnetic fields traced by the 450\,$\rm \upmu m$ and 850\,$\rm \upmu m$ observations correspond to slightly different regions along the LOS, we apply our DCF calculations to explore the relationship between magnetic field strength and density.
In Fig. \ref{fig:B_n_relation}, we show the relationship between the total magnetic field strength and the volume density of molecular hydrogen based on our calculations, comparing with the expectations from \citet{2010ApJ...725..466C} depicted by a dashed line.
We also plot the data for another molecular cloud within the CMZ, G0.253+0.016, obtained from \citet{2016ApJ...832..143F} with DCF method. 
All data points in Fig. \ref{fig:B_n_relation} are positioned above the dashed line.
We fit all data points with Eq. \eqref{eqn:Crutcher_model} excluding the two points of CND because the CND is a torus-like structure rotating around Sgr\,A*, as opposed to being a self-gravitating cloud.
We obtain the power-law index $\kappa=0.21\pm 0.16$ (dash-dotted line in Fig. \ref{fig:B_n_relation}), a value lower than the \citet{2010ApJ...725..466C} result of 0.65.
\citet{2015MNRAS.451.4384T} demonstrated that the power-law index of the B-n relation could be directly related to the cloud geometry and the respective dominant direction of collapse, assuming a frozen-in magnetic field in the absence of significant ambipolar diffusion.
Our result of low $\kappa$ implies that the contraction motions tend to align with the magnetic field lines without significantly altering the magnetic field strengths.
It is important to note, however, that the prescription of the B-n relation is based on spherically collapsing clouds.
Such a framework may not be broadly applicable to such a complex and dynamic region as the inner CMZ.

Additionally, the total magnetic field strength $B_{tot}$ is $\sim611\,\upmu G$ at critical density $n_{0,H_2}=150\,cm^{-3}$, a value roughly 50 times that found in the model by \citet{2010ApJ...725..466C}.
Note, however, that they used the LOS magnetic field strengths to find the relation. 
$B_{LOS}$ is roughly half of $B_{tot}$.
This difference implies that the magnetic fields in the molecular clouds in the CMZ environment are typically stronger than those of the closer regions described in \citet{2010ApJ...725..466C}.

\section{Conclusion} \label{sec:conclusion}
We present the first-look results of the continuum and polarization observations at 450\,$\rm \upmu$m and 850\,$\rm \upmu$m toward the inner region of the Central Molecular Zone using SCUBA-2 with the polarimeter POL-2 as part of the BISTRO project.
Our investigation focuses on the magnetic field morphology and strength in the inner region of the CMZ, covering three distinct areas: the 20MC, 50MC, and CND. 
The main findings of this work are as follows:
\begin{enumerate}
    \itemsep=0.pt
    \item 
    The magnetic field morphology is highly ordered, with the angular dispersion ranging from 5$\degree$ to 10$\degree$.
    In the vicinity of CND, the magnetic field exhibits a spiral structure at both 450\,$\rm \upmu m$ and 850\,$\rm \upmu m$.
    \item 
    We derive average POS magnetic field strengths of 1.0$\rm\pm0.5$\,mG, 1.0$\rm\pm0.5$\,mG, 2.2$\rm\pm1.2$\,mG at 450\,$\rm \upmu m$ and 0.7$\rm\pm0.4$\,mG, 0.8$\rm\pm0.4$\,mG, 1.9$\rm\pm1.1$\,mG at 850\,$\rm \upmu m$ for the 20MC, 50MC, and CND, respectively.
    \item
    Within the 20MC, 50MC and CND, magnetic turbulence is sub-Alfvénic, and the magnetic fields are more dominant than thermal pressure. 
    Additionally, the mass-to-flux ratios are all less than 1, indicating the magnetic field can support these structures against gravitational collapse.
    \item Except for the 450\,$\rm \upmu m$ polarization in the vicinity of the CND, which is affected by poor observation quality, the power-law indices for the dependence of polarization fraction on total intensity range from 0.4 to 0.6, indicating that dust grains are moderately aligned with the magnetic field in the vicinity of the 20MC, 50MC, and the CND.
    \item Previous studies suggest that the low SFR within the CMZ is possibly due to strong turbulence. Our results, however, reveal the presence of a very strong magnetic field, which may be the instead dominant factor to the energy budgets of these clouds.
\end{enumerate}

\section{Acknowledgments}
The James Clerk Maxwell Telescope is operated by the East
Asian Observatory on behalf of the National Astronomical Observatory of Japan; Academia Sinica Institute of Astronomy and Astrophysics; the Korea Astronomy and Space Science Institute; the Operation, Maintenance and Upgrading Fund for Astronomical Telescopes and Facility Instruments, funded by the Ministry of Finance (MOF) of China and administrated by the Chinese Academy of Sciences (CAS), as well as the National Key R\&D Program of China (No. 2017YFA0402700). 
Additional funding support is provided by the Science and Technology Facilities Council of the United Kingdom and participating universities in the United Kingdom and Canada. 
Funding for the construction of SCUBA-2 and POL-2 was provided by the Canada Foundation for Innovation.
The Starlink software \citep[][]{2014StarC..21.....T} is supported by the East Asian Observatory, JCMT Project Code M16AL004. 
This research used the facilities of the Canadian Astronomy Data Centre operated by the National Research Council of Canada with the support of the Canadian Space Agency. 
This research utilized the NASA Astrophysics Data System.
The authors wish to recognize and acknowledge the very significant cultural role and reverence that the summit of Maunakea has always had within the indigenous Hawaiian community. 
We are privileged to conduct observations from this mountain.
We thank the anonymous referee for their thoughtful comments and suggestions, which have helped improve this manuscript.
The data used in this paper were obtained under project codes M20AL018, M17AP074.
M.Z.Y. and S.P.L. acknowledge grants from the National Science and Technology Council (NSTC) of Taiwan 109-2112-M-007-010-MY3 and 112-2112-M-007-011.
This work used high-performance computing facilities operated by the Center for Informatics and Computation in Astronomy (CICA) at National Tsing Hua University. 
This equipment was funded by the Ministry of Education of Taiwan, the National Science and Technology Council of Taiwan, and National Tsing Hua University.
F.P. acknowledges support from the MICINN under grant numbers PID2022-141915NB-C21.
M.T. is supported by JSPS KAKENHI grant No.24H00242. 
J.K. (Jungmi Kwon) is supported by JSPS KAKENHI grant No.24K07086.
W.K. was supported by the National Research Foundation of Korea (NRF) grant funded by the Korea government (MSIT) (RS-2024-00342488).
K.P. is a Royal Society University Research Fellow, supported by grant number URF\textbackslash R1\textbackslash 211322.
X.L. acknowledges support from the National Key R\&D Program of China (No.\textbackslash 2022YFA1603101), the Strategic Priority Research Program of the Chinese Academy of Sciences (CAS) Grant No.\textbackslash XDB0800300, the National Natural Science Foundation of China (NSFC) through grant Nos.\textbackslash 12273090 and 12322305, the Natural Science Foundation of Shanghai (No.\textbackslash 23ZR1482100), and the CAS ``Light of West China'' Program No.\textbackslash xbzg-zdsys-202212.
J.K. (Janik 
Karoly) is supported by the Royal Society under grant number RF\textbackslash ERE \textbackslash 231132, as part of project URF\textbackslash R1\textbackslash 211322.



%

\vspace{5mm}
\facilities{James Clerk Maxwell Telescopy (JCMT)}



%

\software{
    Starlink
    \citep{2014ASPC..485..391C}
    astropy \citep{2013A&A...558A..33A,2018AJ....156..123A,2022ApJ...935..167A}
    }




\bibliography{references}{}
\bibliographystyle{aasjournal}



\appendix

\section{CO Contamination Removal and Missing Flux Issue}\label{appendix:CO_removal}

We attempt to remove the CO emission using the JCMT CHIMPS2 observations from the BISTRO continuum observations during the SCUBA-2 data reduction following the method outlined by \citet{2015MNRAS.453.4264M}.
They treated the CO map as a negative fake source and incorporated it into the reduction pipeline.
Our results, however, suggest that the CHIMPS2 observations are not appropriate for incorporation into our intensity map due to the differences in the scales of the structures traced.

\begin{figure*} [ht!]
\epsscale{0.9}
\plotone{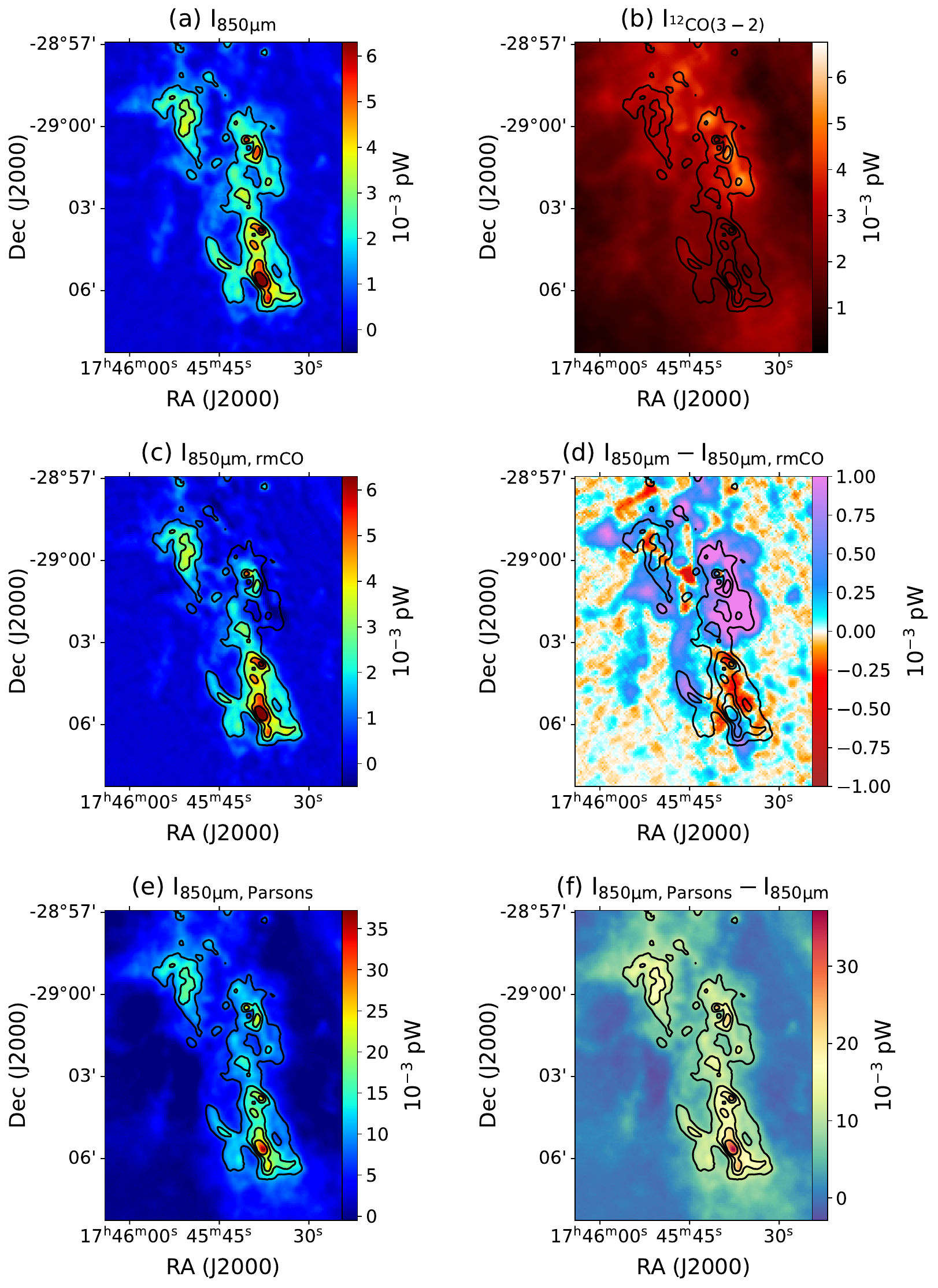}
    \caption{Results of CO contamination removal during the map-making data reduction process.
    Panel (a): original continuum intensity at 850\,$\rm \upmu m$ observed by SCUBA-2 equipped with POL-2.
    Panel (b): $^{12}CO$ (3-2) intensity.
    Panel (c): the intensity at 850\,$\rm \upmu m$ with CO contamination removal during the map-making process using the gas intensity shown in panel (b).
    Panel (d): the gas intensity eliminated from the continuum observations at 850\,$\rm \upmu m$.
    Panel (e): the intensity at 850\,$\rm \upmu m$ from \citet{2018ApJS..234...22P} observed by SCUBA-2.
    Panel (f): Difference between the intensity maps from this work and that of \citet{2018ApJS..234...22P}.
    \label{fig:apx_mapmaking}}
\end{figure*}

First, we calculate the 0th moment map using the $\rm ^{12}CO$ (3-2) observations and convert the units from brightness temperature to pW, the SCUBA-2 output power units, using the parameters in \citet{2015MNRAS.453.4264M}. 
The details are as follows: (1) the 0th moment map is divided by the forward spillover and scattering efficiency $\rm \eta_{fss}=0.71$, (2) multiplied by the conversion factor $C=0.77$ (from effective brightness temperature to $\rm mJy\,beam^{-1}$), and (3) the FCF value of 516 (from $\rm mJy\,beam^{-1}$ to pW) is applied along with the POL-2 transmission factor of 1.35.
We show the BISTRO 850\,$\rm \upmu m$ intensity map and $^{12}CO$ (3-2) intensity map with unit of pW in panel (a) and (b) of Fig. \ref{fig:apx_mapmaking}.
In panel (a), the brightest dust emission at 850\,$\rm \upmu m$ is observed in the 20MC. 
In contrast, the brightest $^{12}$CO (3-2) emission is found near the CND, as shown in panel (b).

Subsequently, we treat the CO intensity map as the fake source map applied to the map-making process.
The resulting dust intensity map with CO contamination removed is shown in panel (c) of Fig. \ref{fig:apx_mapmaking}.
Then, we subtract the CO-removed intensity from the original data and display the result in panel (d).
Unexpectedly, the CO-removed intensity map has larger values than the original maps of the 20MC and 50MC.
We suspect this issue is due to the problem of flux missing from the extended structures.

\citet{2018ApJS..234...22P} produced a CO contamination map using the intensity observations taken by SCUBA-2 without POL-2, in combination with CHIMPS2 $\rm ^{12}CO$ (3-2) data. 
We compare the 850\,$\rm \upmu$m intensity observations from \citet{2018ApJS..234...22P} to our data.
The difference is shown in panel (f) of Fig. \ref{fig:apx_mapmaking}.
This map shows that the intensity map from \citet{2018ApJS..234...22P} captures the larger structure.
The discrepancy is likely due to the different SCUBA-2 observing modes: ``DAISY'' and ``PONG.'' 
The observing mode used in our BISTRO data is DAISY, which is designed for small, compact sources of around 3$'$ or less. In contrast, \citet{2018ApJS..234...22P} utilized the PONG mode, another scan strategy intended for covering large areas \citep[for a detailed intruduction of two observing modes, see][]{2014StarC..21.....T}. 
Additionally, the CHIMPS2 observations were taken with HARP on JCMT, designed to rapidly scan large regions.
The PONG intensity map and the gas emission map obtained with HARP trace structures on similar scales, enabling \citet{2018ApJS..234...22P} to produce a reliable map of CO contamination levels.
In our case, however, the CHIMPS2 observations are not suitable for incorporation into our DAISY intensity map for CO contamination removal.

For results, we adopt the CO contamination fraction derived by \citet{2018ApJS..234...22P} rather than eliminate CO emission during the reduction process, as shown in Fig. \ref{fig:CO_Gas_Fraction}.
The $\rm ^{12}CO$(3-2) contamination ranges from 0\% to 30\% in the vicinity of the 20MC and 50MC, and reaches up to $\sim$80\% in the CND region.
Under the assumption of local thermodynamic equilibrium, the $\rm ^{12}CO$(6-5) contamination is nearly zero in the 20MC and 50MC, but still exceeds $\sim$80\% in the CND region.

\end{document}